\documentclass[twocolumn]{aastex63}

\usepackage{color}

\newcommand\swift{{\it Swift}}
\newcommand\chandra{{\it Chandra}}
\newcommand\sax{{\it BeppoSAX}}
\newcommand\suzaku{{\it Suzaku}}
\newcommand\xmm{{\it XMM-Newton}}
\newcommand\asca{{\it ASCA}}
\newcommand\rxte{{\it RXTE}}
\newcommand\xte{{\it RXTE}}
\newcommand\astrosat{{\it AstroSat}}
\newcommand\nustar{{\it NuSTAR}}
\newcommand\iue{{\it IUE}}
\newcommand\s{{\rm~s}}

\newcommand\mpc{{\rm~Mpc}}

\newcommand\kev{{\rm~keV}}
\newcommand\keV{{\rm~keV}}
\newcommand\ev{{\rm~eV}}
\newcommand\ergs{{\rm~ergs}}
\newcommand\cm{{\rm~cm}}
\newcommand\angstrom{{\rm~\AA}}

\shortauthors{Tripathi et al.}

\begin{document}

\title{Revealing Thermal Comptonization of accretion-disk photons in IC~4329A with \astrosat{}}

\correspondingauthor{P. Tripathi}
\email{prakasht@iucaa.in}


\author[0000-0003-4659-7984]{Prakash Tripathi}
\affiliation{Inter University Centre for Astronomy and Astrophysics, Pune, India, 411007}

\author{Gulab Chand Dewangan}
\affiliation{Inter University Centre for Astronomy and Astrophysics, Pune, India, 411007}

\author{I. E. Papadakis}
\affiliation{Department of Physics and Institute of Theoretical and Computational Physics, University of Crete, 71003 Heraklion, Greece}
 \affiliation{Institute of Astrophysics - FORTH, N. Plastira 100, 70013 Vassilika Vouton, Greece}

\author{K. P. Singh}
\affiliation{Indian Institute of Science Education and Research Mohali, Knowledge City, Sector 81,
Manauli P.O., SAS Nagar, 140306, Punjab, India}
\affiliation{Department of Astronomy and Astrophysics, Tata Institute of Fundamental Research, 1 Homi Bhabha Road, Mumbai 400005, India}

\begin{abstract}
We present five simultaneous UV/X-ray observations of IC~4329A by \astrosat{}, performed over {a five-month} period. We utilize the excellent spatial resolution of the Ultra-Violet Imaging Telescope (UVIT) onboard \astrosat{} to reliably separate the intrinsic AGN flux from the host galaxy emission and to correct for the Galactic and internal reddening, as well as the contribution from the narrow and broad-line regions. We detect large-amplitude UV variability, which is unusual for a large black hole mass AGN, like IC~4329A, over such a small period. In fact, the fractional variability amplitude is larger in the UV band than in the X--ray band. This demonstrates that the observed UV variability is intrinsic to the disk, and is not due to X--ray illumination. The joint X-ray spectral analyses of five SXT and LAXPC spectral data reveal a soft-X-ray excess component, a narrow iron-line (with no indication of a significant Compton hump), and a steepening power-law ($\Delta\Gamma\sim 0.21$) with increasing X-ray flux. The soft excess component could arise due to thermal Comptonization of the inner disk photons in a warm corona with $kT_e\sim 0.26$ keV. The UV emission we detect acts as the primary seed photons for the hot corona, which produces the broadband X--ray continuum. The X-ray spectral variability is well described by the cooling of this corona from $kT_e\sim42\kev$ to $\sim 32\kev$ with increasing UV flux, while the optical depth remains constant at $\tau\sim 2.3$.

\end{abstract}

\section{Introduction}  \label{sec:intro}

The primary emission from type 1 active galactic nuclei (AGN) consists of the big blue bump (BBB) in the $1~\mu$m to $1000$~\AA\,  range,  a soft X-ray excess component below $\sim 2\kev$, and a broadband X-ray power-law continuum, with a high energy cut-off at a few $100\kev$ \citep{2015MNRAS.451.4375F,2016MNRAS.458.2454L, 2017ApJS..233...17R,2018A&A...614A..37T}. Additional spectral components such as the infra-red bump, the optical/UV emission lines, the X-ray reflection (consisting of iron K$\alpha$ line, and the Compton reflection hump in the  $20-40\kev$ band), and at least a fraction of the soft excess emission can arise due to the reprocessing of the primary emission. 

The central engine in AGN responsible for the primary emission is thought to consist of an accretion disk surrounding a supermassive black hole, and a compact hot corona ($kT_e \sim 100\kev$) in the innermost regions. The BBB component is believed to be the direct signature of the accretion flow and is interpreted as thermal emission from the accretion disk. This component is thought to interact with the hot corona. Photons emitted by the disk undergo repeated Compton up-scattering by energetic electrons in the corona, thus  giving rise to the broadband primary X-ray power-law like emission \citep{1993ApJ...413..507H}. The X-ray power-law in turn illuminates the accretion disk that can absorb and reflect the X-rays, giving rise to the broad iron line, reflection hump and at least some fraction of the soft X-ray excess emission. The soft excess emission can also be produced by thermal Comptonization of the optical/UV disk photons by a warm ($kT_e\sim 0.3\kev$) and optically thick layer of inner disk below a few tens of { gravitational radii} \citep{2007ApJ...671.1284D,2015A&A...575A..22M, 10.1111/j.1365-2966.2011.19779.x, 2018MNRAS.480.1247K,2018A&A...611A..59P,2020A&A...634A..85P} though it is not clear if such a warm corona can be produced in the inner disk. 

Observations of the broadband X-ray power-law emission cutting-off at high energies in a number of AGN {provide} strong support to the thermal Comptonization model. However, the nature of the seed photons still remains to be explored. If the variability of the Comptonized X-rays were to be dominated by the variations in the seed flux, it would have been easier to identify the seed photons. But this is not the case, usually. In most cases the X--ray variability amplitude is significantly larger than the variability amplitude we observe in the optical/UV bands. In addition, observations of time lags in the optical/UV variations relative to those in the X-rays imply that it is the X-ray emission that drives the optical/UV variability. Such optical/UV lags are found to be wavelength-dependent and easily explained as arising due to the X-ray reprocessing by the accretion disk~\citep{Sergeev_2005,10.1111/j.1365-2966.2007.12098.x,10.1111/j.1365-2966.2012.20677.x,2016MNRAS.456.4040T,2018MNRAS.480.2881M,2019ApJ...870..123E,2020ApJ...896....1C}. To date, there are only a few observations of optical/UV emission leading X-rays that are consistent with Comptonization delays~\citep{Adegoke_2019}. 

In the thermal Comptonization model,  variations in the seed flux are expected to affect both the X--ray flux and spectral variability in AGN. A number of Seyfert 1 AGN show X-ray spectral steepening with increasing X-ray flux. Such spectral variability can be caused by cooling of the corona  due to increased irradiating seed flux. For example, based on a month long monitoring observation of NGC~7469 simultaneously with \iue{} and \xte{}, \cite{Nandra_2000} found that the X-ray  photon-index is correlated with the intrinsic UV flux and suggested thermal Comptonization of UV photons as the origin of the X-ray continuum. \cite{Zdziarski_2001} showed that the observed softening of X-ray emission in 3C~120 from $\Gamma\sim1.7$ to $\sim 2$ with increasing soft X-ray flux could be explained in terms of thermal Comptonization by requiring the irradiating optical/UV flux to increase by a factor of $\sim 2$. However, this model could not be tested fully in the absence of simultaneous optical/UV observations.  \cite{2013MNRAS.433.1709G}  reported a correlation between the photon-index of the power-law component and the soft X-ray ($0.1-1\kev$) flux in an X-ray-bright radio-loud NLS1 galaxy PKS~0558--504. Here, we use \astrosat's multi-wavelength capability for simultaneous observations in the near and far UV, soft and hard X-rays, and investigate the UV/X-ray connections in IC~4329A. 

IC~4329A is a nearby AGN at a redshift of $z = 0.016054$~\citep{1991AJ....101...57W} and the second brightest type 1 AGN in the \swift{}/BAT catalog \citep{2013ApJS..207...19B}. The AGN has been classified as Seyfert 1.2~\citep{2006A&A...455..773V} residing at the center of an edge-on host galaxy with a dust-lane passing through the nucleus. IC~4329A has been studied with all major X-ray satellites including 
\asca{}/\rxte{}~\citep{2000ApJ...536..213D}, \sax{}~\citep{2002A&A...389..802P}, \chandra{}~\citep{2004ApJ...608..157M}, \xmm{}~\citep{2007MNRAS.382..194N}, \suzaku{}~\citep{2016MNRAS.458.4198M}, and jointly with \suzaku{} \& \nustar{}~\citep{2014ApJ...788...61B}. These observations suggest a modest or weak broad iron line that does not require X-ray reflection from the inner disk. \cite{2018A&A...619A..20M} derived the broadband spectral energy distribution of IC~4329A using the UV and optical observations with \swift{}, high resolution \xmm{} and \chandra{} X-ray data, and mid-IR spectroscopy, and estimated an accretion rate of $10-20\%$ of the Eddington limit for a given black hole mass in the range of $1-2\times10^8M_{\odot}$. 

{ We observed IC~4329A five times with the Indian  multi-wavelength astronomy  satellite  \astrosat{} \citep{2014SPIE.9144E..1SS, 2006AdSpR..38.2989A} in the period between February to June 2017. In a companion  paper, \citet{Dewangan_et_al_2021} studied in detail the Ultra-Violet Imaging Telescope (UVIT) data from these observations. They computed the mean source flux in the FUV (F154W, $\lambda_{mean}=1541$\angstrom{}, $\Delta\lambda=380$\angstrom{}), and NUV (N245M, $\lambda_{mean}=2447$\angstrom{}, $\Delta\lambda=280$\angstrom{}) filters, and found that the source 
is not detected in the FUV band. Using the UVIT and archival {\it HST} data, they found that the 
intrinsic UV continuum of the active nucleus is fully consistent with standard accretion disk models, 
but only if the disk emits from distances larger than 80--150 gravitational radii (depending on the assumed extinction law). In this work we focus on the analysis of the X-ray data from the same observations, and the study of the relation between the observed UV and X-ray variations.}

This paper is organized as follows. We first describe observations and the data reduction in section~\ref{obs}. We describe in detail the UVIT data analysis in section~\ref{uvitanalysis}. We present the results from the spectral analysis in section~\ref{analysis}. We discuss implications of our results in section~\ref{discussion}, and we conclude in section~\ref{conclusion}.

\section{\astrosat{} observations and Data reduction} 
\label{obs}

Details of the \astrosat{} observations are listed in Table~\ref{tab:1}. \astrosat{} carries four co-aligned instruments -- the Ultra-Violet Imaging Telescope (UVIT; \citealt{Tandon_2017, 2020AJ....159..158T}), the  Soft X-ray Telescope (SXT; \citealt{10.1117/12.2235309, 2017JApA...38...29S}), the Large Area X-ray Proportional Counter (LAXPC; \citealt{2016SPIE.9905E..1DY, 2017JApA...38...30A, Antia_2017}), and the Cadmium-Zinc-Telluride Imager (CZTI; \citealt{2016SPIE.9905E..1GV}). We used the data acquired  with the UVIT, SXT and LAXPC. We did not use CZTI data as the source was not detected. 

\subsection{The SXT Data}

SXT is a focusing X-ray telescope with a CCD camera similar to those employed by the \swift{}/XRT and \xmm{}/MOS. It operates in the photon counting mode and is capable of low resolution  imaging  (Full Width at Half Maximum (FWHM) $\sim 2{\rm~arcmin}$, Half-Power Diameter (HPD) $\sim 11{\rm~arcmin}$) and medium energy resolution  spectroscopy (FWHM $\sim 150\ev$ at $6\kev$) in the $0.3-8\kev$ band. We processed the level-1 SXT data with the latest pipeline software AS1SXTLevel2-1.4b, which is available at the SXT payload operation centre (POC) website\footnote{\url{https://www.tifr.res.in/~astrosat_sxt/sxtpipeline.html}}, and we generated the level-2 clean event files for individual orbits. We merged the clean event files corresponding to a  given observation id using the {\it Julia} SXT event merger tool {\textsc{sxt\_l2evtlist\_merge}}. This tool is developed by us and is available at the SXT POC website. It identifies common intervals in the event files and retains only unique events.  

\begin{figure*}
    \centering
    \includegraphics[scale=0.35]{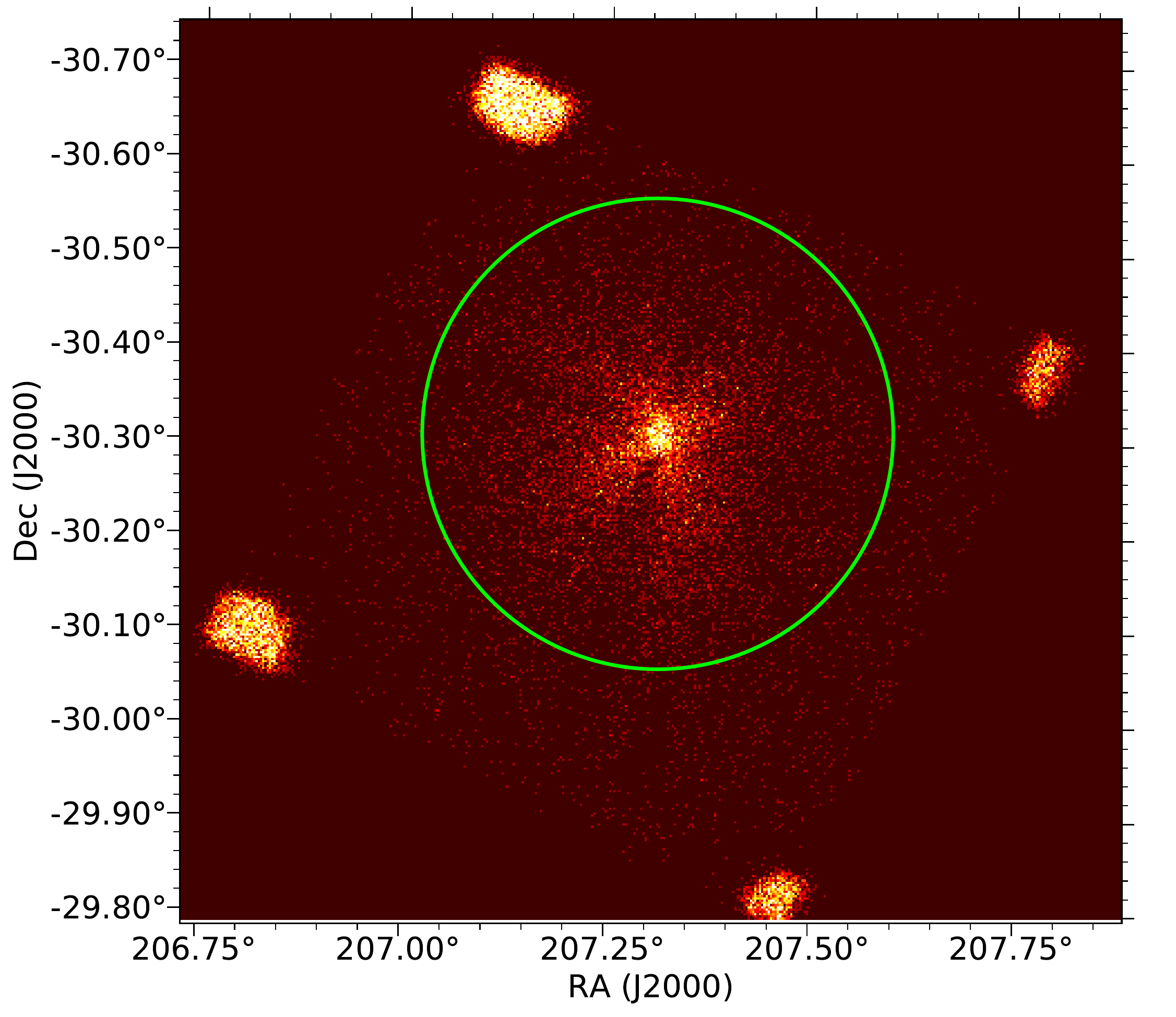}
    \includegraphics[scale=0.35]{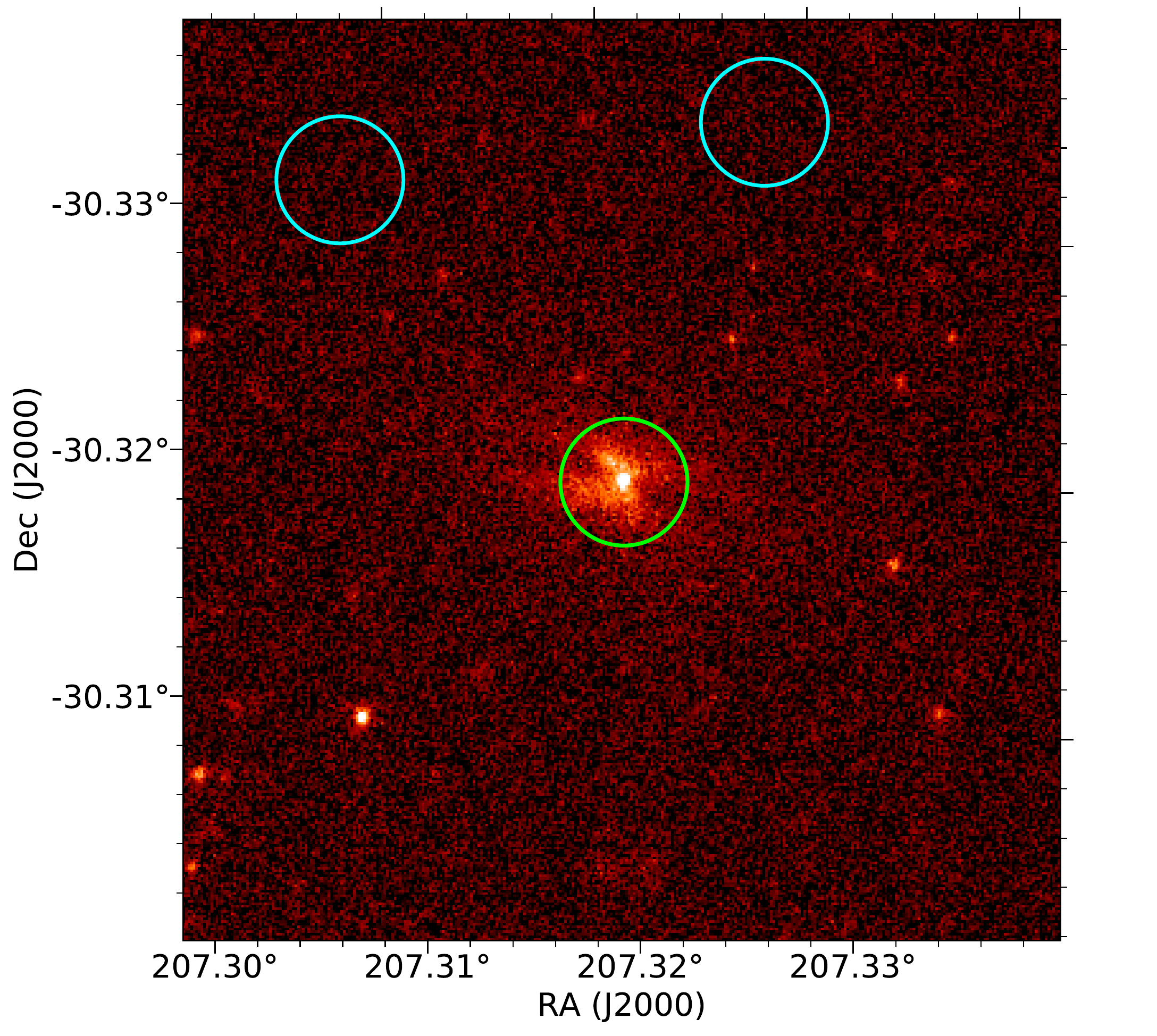}
    \caption{X-ray ({\it left}) and NUV ({\it right}) images of IC~4329A for observation 9000001048, obtained from SXT and UVIT data. The NUV image shows the galaxy and the dust-lane in the central region ($2.4' \times 2.5'$). Green circuls in both images indicate the source counts extraction regions, and the two cyan circles in the NUV image represent the background extraction regions.}
    \label{figimages}
\end{figure*}

{ The SXT image of IC~4329A from observation 9000001048 is shown in the left panel of Figure~\ref{figimages}}. 
We used the {\textsc{xselect}} tool available within HEASoft\footnote{\url{https://heasarc.gsfc.nasa.gov/docs/software/lheasoft/}} version 6.26.1 to extract the source spectrum. For each observation, we extracted the spectrum from the merged event list using a circular region of 15$\arcmin$ radius centered on the source (indicated by the green circle in Figure~\ref{figimages} ({\it Left}). The large HPD and the four corner calibration sources leave the SXT CCD camera with virtually no source-free regions.  Hence, we used the blank sky background spectrum ({\fontfamily{qcr}{\textrm{SkyBkg\_comb\_EL3p5\_Cl\_Rd16p0\_v01.pha}}}) provided by the SXT instrument team. We also used the most recent spectral  redistribution matrix file ({\fontfamily{qcr}{\textrm{sxt\_pc\_mat\_g0to12.rmf}}}) and ancillary response file ({\fontfamily{qcr}{\textrm{sxt\_pc\_excl00 v04\_20190608.arf}}}) available at the SXT POC webpage. We grouped each spectral dataset using HEASoft task {\textsc{grppha}} to have a minimum of 30 counts per bin. The net 0.3--7.4 keV count rates are listed in the fourth column of Table \ref{tab:1}. The count rate varies by more than a factor of 2, from $0.74{\rm~counts~s^{-1}}$ to $1.67{\rm~counts~s^{-1}}$ during the five observations. 

\begin{table*}
	\caption{\astrosat{} observations of IC 4329A.}
	\label{tab:1}
	\centering
	\begin{tabular}{ccccccccc}
		\hline\hline
		Obs. ID        & Date       &\multicolumn{2}{c}{SXT} &\multicolumn{2}{c}{LAXPC} &\multicolumn{3}{c}{N245M/UVIT}\\	
		&yyyy-mm-dd	& Exposure	&Rate$^a$  & Exposure 	&Rate$^b$ 			& Exposure 	&Net source rate$^{c}$ &bkg rate$^{d}$ \\
		&			&(ksec)		& (cts s$^{-1}$)	& (ksec)	&(cts s$^{-1}$)	& (ksec)	&(cts s$^{-1}$)&(cts s$^{-1}$)\\
		\hline
		
		9000001006    &2017-02-03	 &26.0       & 1.38 $\pm$ 0.008 &44.7  &14.04 $\pm$ 0.030 	&17.26	&0.499 $\pm$ 0.007	&0.175 $\pm$ 0.003\\
		9000001048    &2017-02-23   &24.5       &1.40 $\pm$ 0.007 	&40.8  &15.37 $\pm$ 0.032   &19.97	&0.529 $\pm$ 0.007&0.170 $\pm$ 0.003\\
		9000001118	  &2017-03-29	 &26.7		  &0.74 $\pm$ 0.006 & 33.0	&08.19 $\pm$ 0.031  &19.63	&0.499 $\pm$ 0.007&0.171 $\pm$ 0.003\\
		9000001286    &2017-06-11   &25.0       &1.62 $\pm$ 0.008 	& 39.0  &13.04 $\pm$ 0.031  &18.83	&0.524 $\pm$ 0.007&0.170 $\pm$ 0.003\\
		9000001340    &2017-06-25   &20.2       &1.67 $\pm$ 0.009  & 41.6  &13.57 $\pm$ 0.031  &18.47	&0.549 $\pm$ 0.007&0.174 $\pm$ 0.003\\
		\hline\hline		  			
	\end{tabular}
\\$^a$ Background-corrected net SXT count rates in $0.3-7.4\kev$ band\\
$^b$ Background corrected net LAXPC count rates in $^b$$4-20\kev$ bands\\
$^c$ Background-corrected net count rates of IC~4329A (AGN+host galaxy) extracted using a circular aperture of 10$\arcsec$ radius.\\
 $^d$ The average background count rate derived from 15 source-free circular regions of 10$\arcsec$ radius. \\

\end{table*}

\subsection{The LAXPC Data}

The LAXPC consists of three identical and co-aligned X-ray proportional counters units (LX10, LX20 and LX30), operating in the $3-80\kev$ band.  The LAXPC operates in the normal mode which is the combination of the Broad Band Counting (BBC) and Event Analysis (EA) modes. We did not use the data acquired with the LX10 and LX30. LX10 is not suitable for faint sources such as AGN due to its unstable nature, while LX30 suffered with continuous gain shift caused by gas leakage \citep{Antia_2017}.  We processed the LX20 data using the pipeline LAXPCSOFT V3.0\footnote{\url{https://www.tifr.res.in/~antia/laxpc.html}} provided by the instrument team. We considered only the layer-1 data to reduce the background and generated the spectrum. The LAXPCSOFT also generates suitable background spectrum using the blank-sky observations performed close to the source observation. We used the response file {\textrm{lx20L1v1.0.rmf}} for our spectral analysis. We grouped the spectrum using {\textsc{grppha}} task to have a minimum of 20 counts per energy bin. The net count rates in LX20 energy band of 4-20 keV are listed in the sixth column in Table \ref{tab:1}. The source cunt rate varied by almost a factor of 2, from $8.2$ to $15.4{\rm~counts~s^{-1}}$ between the five observations. 

\subsection{The UVIT Data}
The UVIT has two telescopes -- one for the far ultraviolet channel (FUV; 1300-1800\AA), and the  other for the near-ultraviolet (NUV; 2000-3000\AA) and visible (VIS; 3200-5500\AA) channels. The UVIT is primarily an imaging telescope with excellent spatial resolution (FWHM$\sim1-1.5{\rm~arcsec}$). Each channel has a number of broadband filters with limited band-passes. The FUV and NUV channels operate in the photon counting mode and are well calibrated for photometric studies. The VIS channel is operated in the integration mode, and it is used for tracking purposes only. We used the F154W (FUV BaF$_2$) and N245M (NUVB13) filters for all the five observations.  We processed the UVIT data of each observation using {\textsc{ccdlab}} \citep{2017PASP..129k5002P}, and generated cleaned images for each observation. The AGN in IC~4329A is not detected in the FUV band due to a dust lane passing through the nucleus \citep{Dewangan_et_al_2021}, therefore we use only the NUV data here. { The central part ($2.4' \times 2.5')$ of the  NUV image of IC~4329A is shown in the right panel of Figure~\ref{figimages}. The bright active nucleus, the diffuse emission from the edge-on host galaxy and a dust lane passing through the central regions are clearly seen in this figure.}

\section{The UVIT data analysis}
\label{uvitanalysis}




As a first test to examine possible variations in the NUV emission from the AGN, we performed aperture photometry of the central region of IC~4329A and a point-like source, (WISEA~J134854.55--302140.7, which is classified as an ultraviolet source in NED\footnote{\url{https://ned.ipac.caltech.edu/}}, { and its flux is relatively constant). Comparison between the light curves of IC~4329A and this source will directly indicate the presence of intrinsic variations in the active nucleus.} We extracted source counts from a circular region with a radius of 25 pixels ($\sim 10\arcsec$), centered on IC~4329A (indicated by the green circle in the right panel of Figure~\ref{figimages}). Background counts were extracted from 15 source-free regions of the same circular size, around IC~4329A (two of them are shown with cyan circles in the right panel of Figure~\ref{figimages}). We averaged the background counts from the 15 source-free regions, and calculated the net source counts by subtracting the averaged background counts from the source counts. We list the net source count rates for the IC~4329A and the averaged background count rates in the last columns of Table~\ref{tab:1}. For the point source, we used  a circular region with a radius of 10 pixels ($\sim 4\arcsec$), centered on the source and a similar size of nearby source-free region to calculate the background.

\begin{figure*}
	\centering
	\includegraphics[scale=0.4]{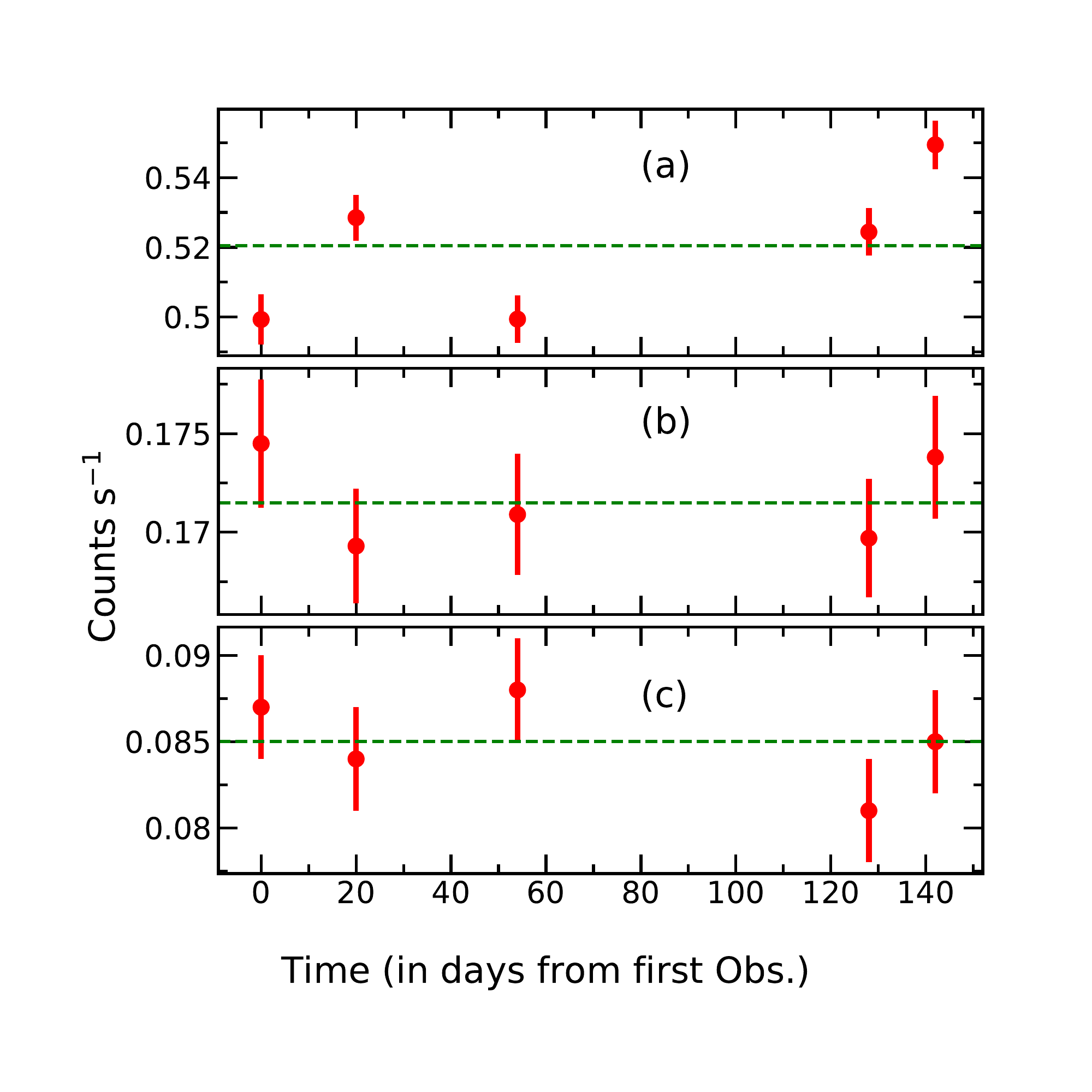}	
	\includegraphics[scale=0.4]{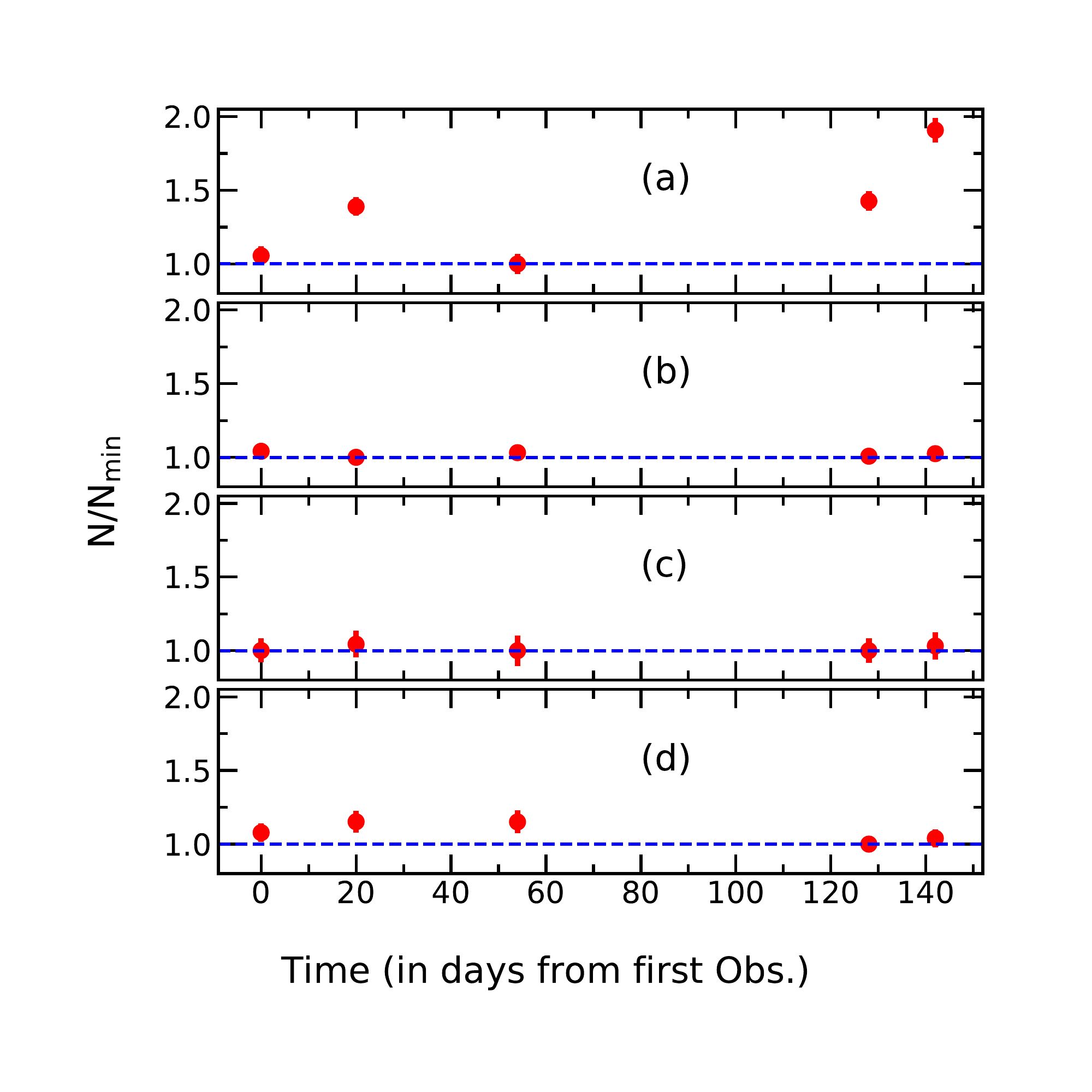}
	\caption{{\it Left:} The NUV  light-curves of (a) IC~4329A (AGN+host galaxy), (b) background, and (c) point source, using count rates derived from aperture photometry (see \S \ref{uvitanalysis}). The green dashed lines in each panel indicate the best-fit constant count rate. {\it Right:} The NUV light-curve of (a) the AGN, (b) background, (c) host galaxy, and (d) point source, using  count rates derived from the radial profile analyses (see, Table~\ref{tab:nuvres}). The light curves are normalized to the minimum count rate (the N/N$_{\rm min}=1$ value is indicated by the blue dashed lines in each panel)}.
	\label{fig:uvlcurve}
\end{figure*}

We plot the light-curves of IC~4329A, the background, and the point-source derived from the aperture photometry in Figure~\ref{fig:uvlcurve} ({\it left panel}). We fitted the light-curves with a constant model that resulted in reduced $\chi^2_{\nu}$ of 9.3, 0.6, and 0.8 for IC~4329A (AGN plus galaxy), background, and point source, respectively. The best-fit constants are shown in the horizontal green dashed lines in Figure~\ref{fig:uvlcurve}. Clearly, the NUV emission from the active nucleus in IC~4329A is highly variable. The constant emission from the point source and the steady background level demonstrate that NUV variability of IC~4329A is real.

\begin{figure*}
    \centering
    \includegraphics[scale=0.35]{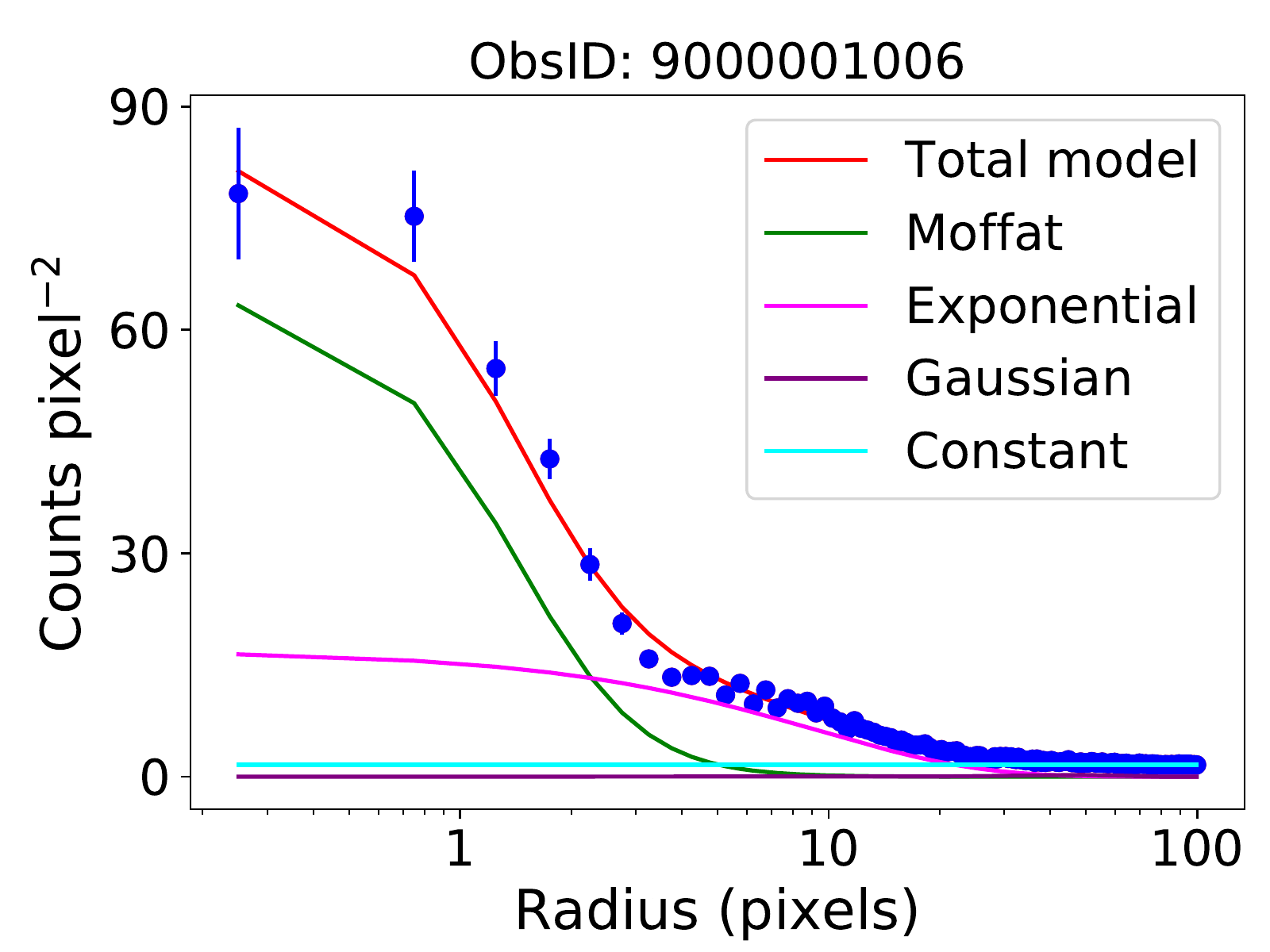}
    \includegraphics[scale=0.35]{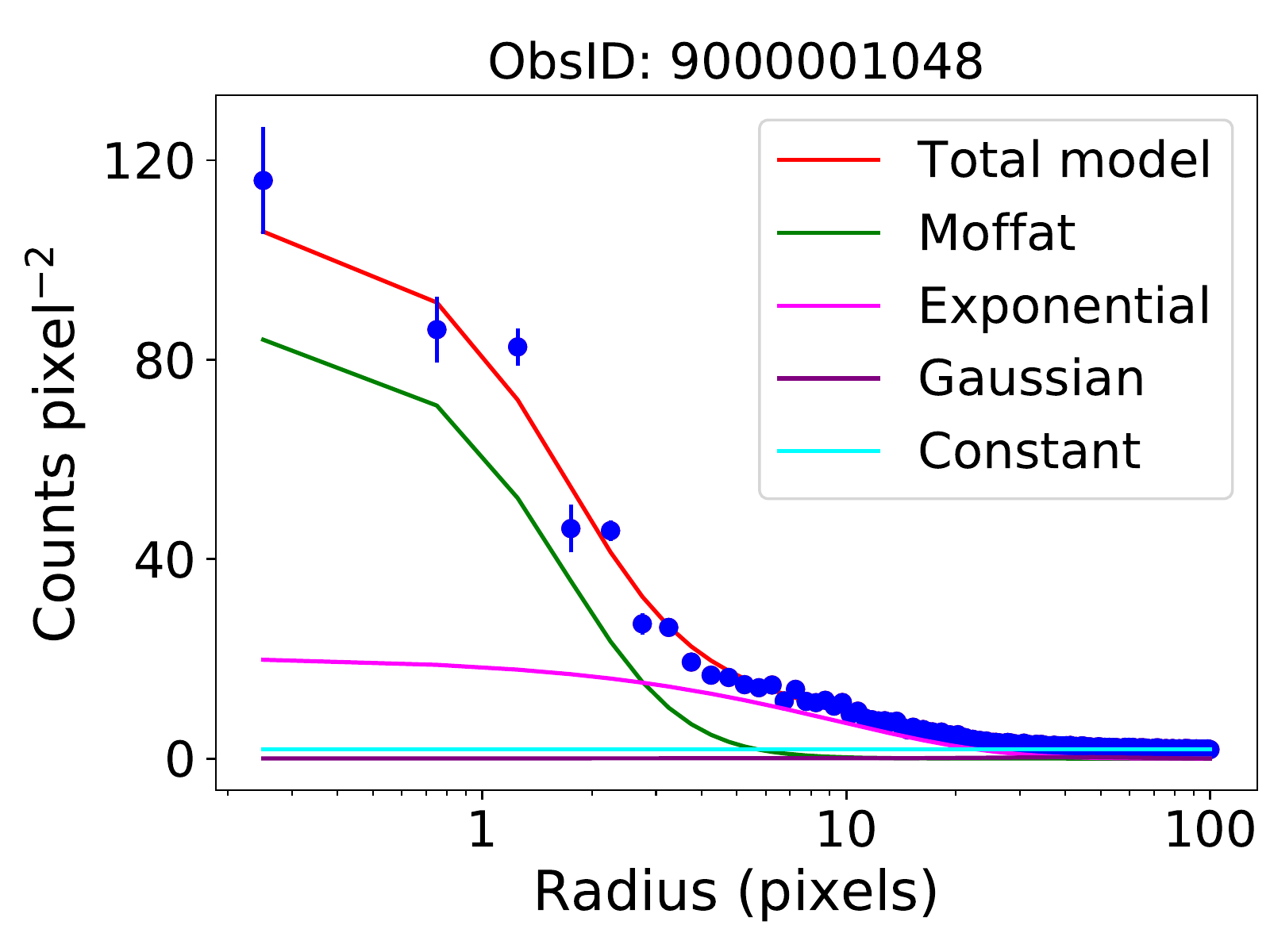}
    \includegraphics[scale=0.35]{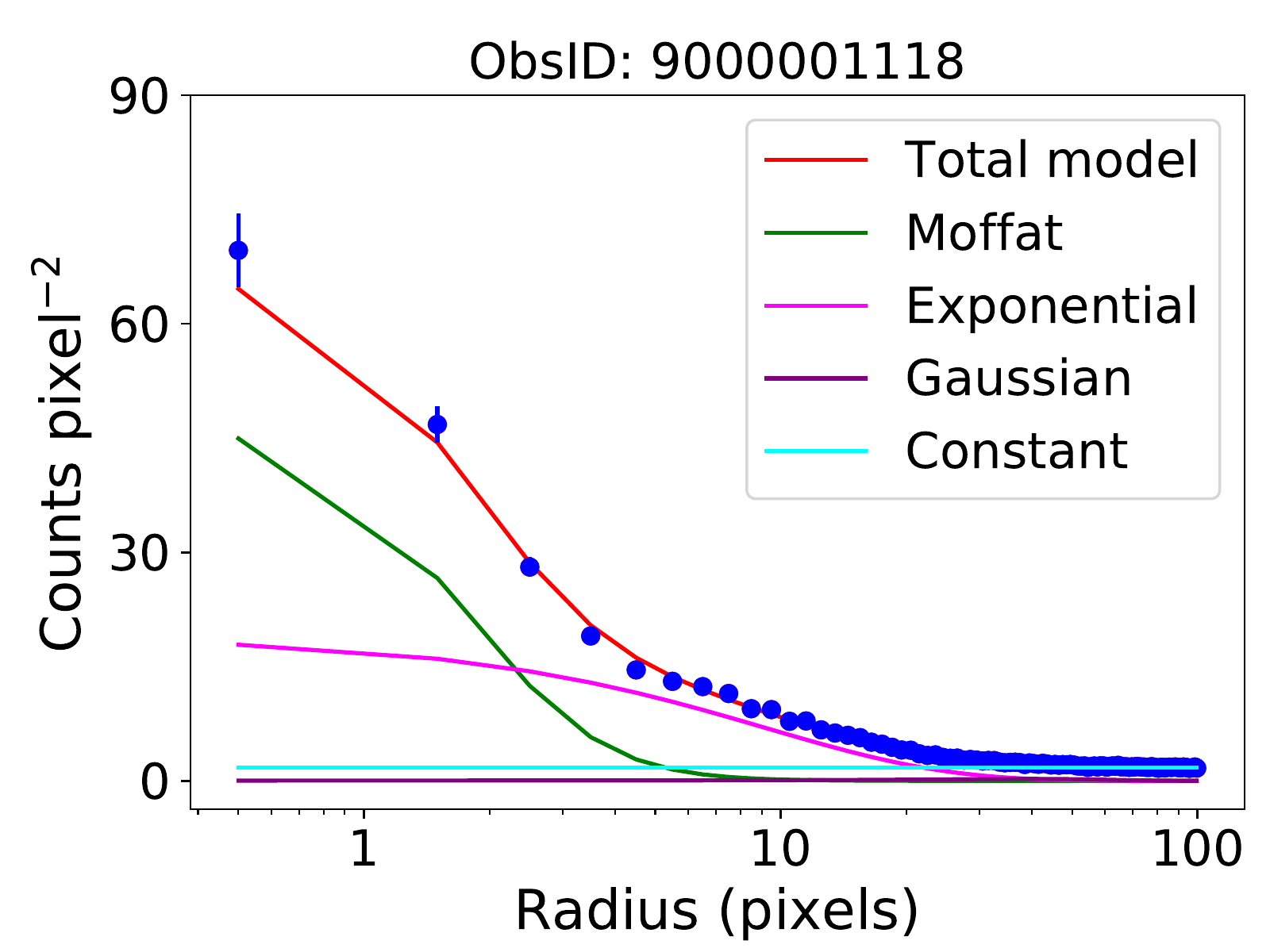}
    \includegraphics[scale=0.35]{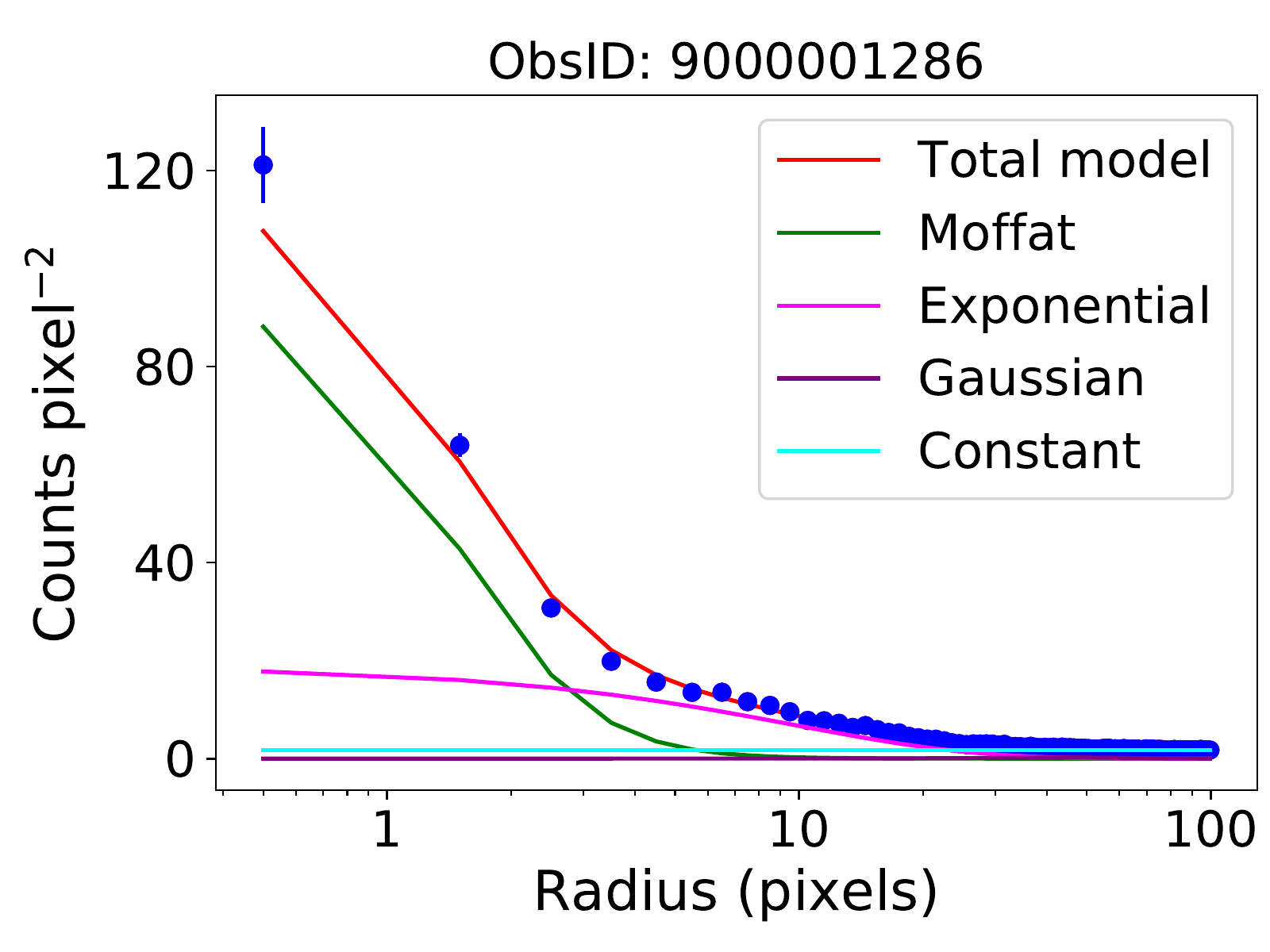}
    \includegraphics[scale=0.35]{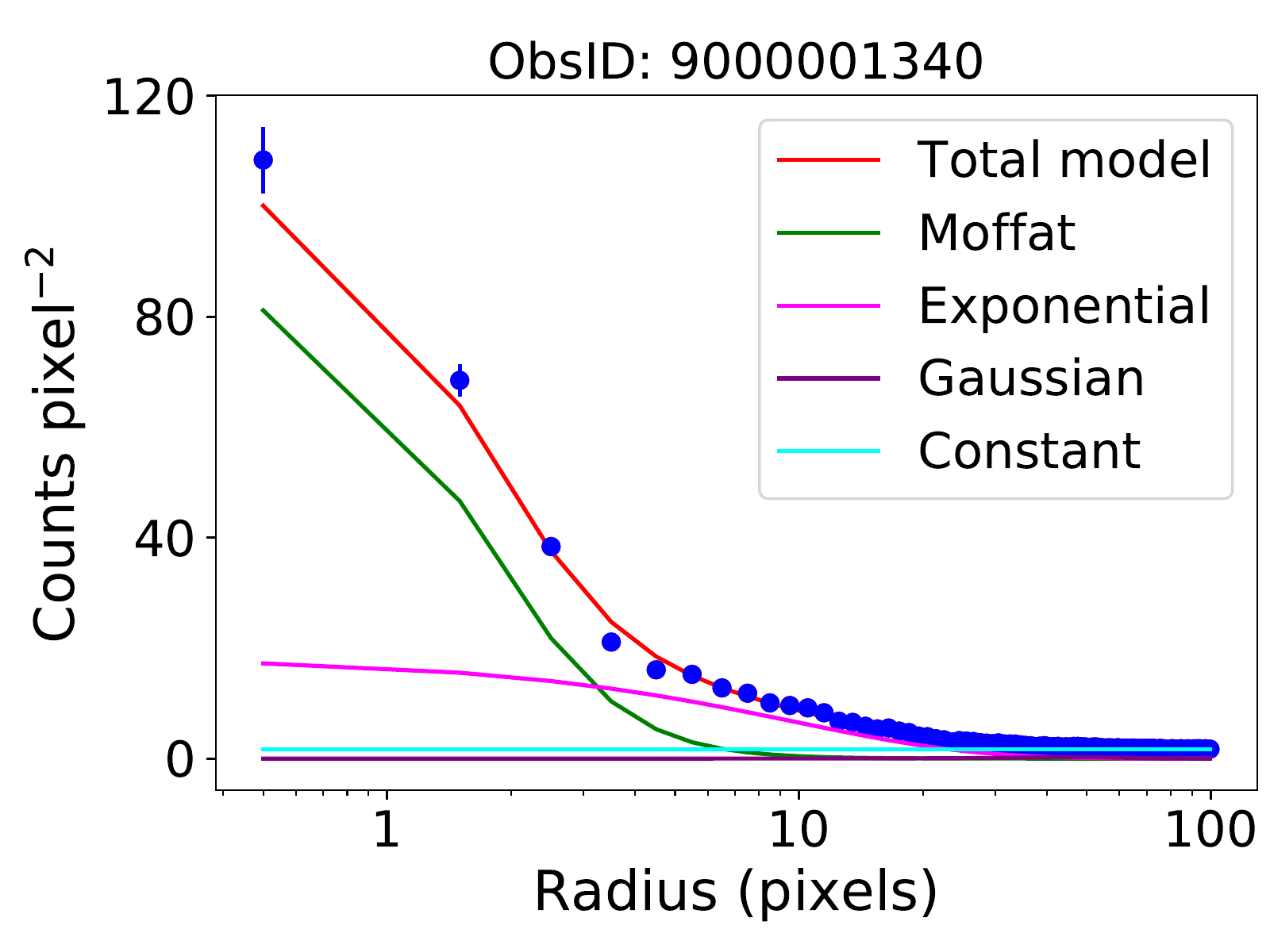}
    \caption{The radial profiles of IC~4329A in all observations, and the best-fit  Moffat function (green), exponential (magenta), and a constant (cyan) components (purple lines indicate the Gaussian component used to account for an emission feature located further away from the galaxy). Red lines show the overall model fits.}
    \label{figradprof}
\end{figure*}

\begin{table*}
	\centering
	\caption{Intrinsic count rates and {flux} of IC~4329A in the N245M/UVIT filter (we list 1-$\sigma$ errors).}
	\label{tab:nuvres}
	\begin{tabular}{c c c c c c c }
		\hline\hline
			Component			&9000001006	&9000001048	&9000001118	&9000001286	&9000001340\\	
		\hline
		Background$^a$	&$1.091\pm0.009$	&$1.048\pm0.011$	&$1.081\pm0.013$	&$1.056\pm0.012$	&$1.074\pm0.009$\\
		\\\
		Galaxy$^a$	&0.692 $\pm$ 0.056	&0.721 $\pm$ 0.063	&0.691 $\pm$ 0.072	&0.691 $\pm$ 0.058	&0.713 $\pm$ 0.063\\
		\\\
		AGN$^a$	&0.057 $\pm$ 0.0035	&0.075 $\pm$ 0.0034	&0.054 $\pm$ 0.0037	&0.077 $\pm$ 0.0036	&0.103 $\pm$ 0.0045\\
		\\\
		Galactic extinction		&0.082 $\pm$ 0.0050	&0.108 $\pm$ 0.0049 &0.077 $\pm$ 0.0053	&0.110 $\pm$ 0.0052&0.147 $\pm$ 0.0057\\
		corrected$^a$			&&&&&&\\
	
		Intrinsic extinction	&35.26 $\pm$ 2.16 &46.49 $\pm$ 2.11	&33.30 $\pm$ 2.29	&47.65 $\pm$ 2.25 &63.49 $\pm$ 2.46\\
		corrected$^a$		&&&&&&\\
		
		Non-continuum &29.02 $\pm$ 1.78	&38.25 $\pm$ 1.74	&27.40 $\pm$ 1.88	&39.21 $\pm$ 1.85 &52.25 $\pm$ 2.03\\
		subtracted$^a$		&&&&&&\\
		Intrinsic flux ($f_{NUV}$)$^b$		&21.04 $\pm$ 1.29	&27.74 $\pm$ 1.27	&19.87 $\pm$ 1.37	&28.43 $\pm$ 1.35	&37.88 $\pm$ 1.48\\
		\hline\hline
	\end{tabular}
	 \\
	$^a$ In cts s$^{-1}$. \\
	$^b$ In units of 10$^{-15}$ $\ergs{}\cm^{-2}\s^{-1}$\angstrom{}$^{-1}$.\\
	
\end{table*}

The UV emission from nearby AGN observed with broadband filters is contaminated by the host galaxy emission, and numerous emission lines from the broad and narrow line regions, including the Fe~II and the Balmer continuum emission. It is also reddened due to the internal and Galactic absorption. Therefore, following \cite{Dewangan_et_al_2021}, first we separated the emission from the AGN and the host galaxy, then we corrected for the Galactic and internal extinction, and we subtracted the emission line emission. We describe the steps below.


\subsection{Radial profile analysis}

We utilized the excellent spatial resolution of the UVIT to reliably separate the host galaxy and the AGN emission by constructing and fitting radial profiles of the source. We first determined the point spread function (PSF) of the instrument. For this purpose, we extracted the radial profile of WISEA~J134854.55--302140.7s,  and we fitted it with a Moffat function ($M(x) = M_0[1 + (\frac{x}{\alpha})^2]^{-\beta}$, FWHM = $2\alpha\sqrt{2^{1/\beta} - 1}$) plus a constant for the background. We treated the best-fitting Moffat profile as the PSF. Analysis of the radial profile of other point sources in the field provided similar best-fit results when fitted by the same function. Tracking correction due to  pointing jitter may lead to slightly different PSFs for different observations. We therefore repeated the above exercise for all five observations and determined the PSF for each observation. The  FWHM of the best-fitting Moffat function is $1.06\arcsec$, $1.24\arcsec$, $1.34\arcsec$, $1.10\arcsec$, and $1.29\arcsec$ in the sequence of observation ids, as listed in Table~\ref{tab:1}. 

For each observation, we fitted the radial profile of IC~4329A with the corresponding PSF, to account for the AGN emission, plus an exponential profile (i.e. $I = I_0\exp(-r/r_d)$) for the host galaxy emission, and a constant component for the background\footnote{We also added a Gaussian to account for an emission feature at a distance larger than $\sim 30$ pixels away from the galaxy. Its amplitude is much smaller than the galaxy of the active nucleus amplitude, but it is comparable to the background, and its addition helps us to determine the background level more accurately. We plot this component in Figure~\ref{figradprof}. Its low amplitude makes it difficult to spot this component and shows that it does not affect the flux measurement of either the galaxy or the active nucleus.}. The best-fit resulted in $\chi^2_{\nu} = 1.18, 0.97, 1.02, 1.29$, and $1.2$ in the sequence of observation ids. The radial profiles of IC~4329A extracted from the five observations and the corresponding best-fit models {along with} the individual components are shown in Figure~\ref{figradprof}. We integrated the best-fit Moffat function, the exponential profile, and the constant component, over a radius of 60 pixels ($\sim 24\arcsec$), and we calculated the model count rates for the AGN, the host galaxy, and the background, respectively. The count rates thus derived are listed in Table~\ref{tab:nuvres} (1st, 2nd, and 3rd rows). 

\subsection{Extinction correction}

We used the extinction curve of \cite{1989ApJ...345..245C}  for the Galactic extinction with a color excess of $E(B-V) = 0.052$ and the ratio of total to selective extinction $R_V = A_V/E(B-V) = 3.1$, 
where $A_V$ is the extinction in the $V$ band. The AGN count rates corrected for the Galactic-extinction are listed in the 4th row in Table~\ref{tab:nuvres}. Following \cite{2018A&A...619A..20M}, we used the extinction curve of \cite{2004MNRAS.348L..54C} with a color excess of $E(B-V) = 1$ for the intrinsic extinction. The AGN count rates corrected for the internal reddening are also listed in Table~\ref{tab:nuvres} (5th row).

\subsection{Correction for the contributions from the broad/narrow line regions}
Finally, we used the composite quasar spectrum  of \cite{2001AJ....122..549V} and calculated the contribution of emission lines including the Fe~II features and the Balmer continuum to the NUV emission from the AGN. We first determined the continuum of the composite quasar spectrum by fitting the $1350-1365$ and $4200-4230$\angstrom{} regions with a power-law model. We then used the effective area of the N245M filter and estimated the NUV count rates of the continuum and the total spectrum. We found that the fractional contribution of the non-continuum components is $\sim17.7\%$ for this filter. Assuming the presence of a similar fraction of the non-continuum components in IC~4329A, we subtracted this contribution from the extinction corrected AGN count rates to compute the intrinsic disk emission (results are listed in the 6th row of Table~\ref{tab:nuvres}). We then converted the intrinsic source count rates to flux density (in $\ergs{}\cm^{2}\s^{-1}$ \angstrom{}$^{-1}$), using the flux conversion factor for the N245M filter \citep{Tandon_2017,2020AJ....159..158T}. The intrinsic fluxes of the source are given in the last row of Table~\ref{tab:nuvres}. 
 
Figure~\ref{fig:uvlcurve} ({\it right panel}) shows the AGN, background, host galaxy and the point source light curves in the NUV/N245M band (from top to bottom, respectively), using the values listed in Table~\ref{tab:nuvres}. The light curves are normalized to the respective minimum count rate. Clearly, the most variable light curve is that of the AGN. The background, host galaxy and the point source emission are almost constant over the same period. This is consistent with the results plotted in the left panel of Figure~\ref{fig:uvlcurve}, and shows that the variations detected in the central region of IC~4329A with the simple aperture photometry are due to the AGN. It also shows that the variations we detect in the AGN emission are genuine, rather than an artefact of our radial profile analysis. We observe significant variations with a factor of $\sim 2$, over a period of just 6 months. This large amplitude variability in the near UV band on such a short timescale is rather remarkable for an AGN with a black-hole mass larger than $10^8$ M$_{\odot}$. 


\section{Spectral Analysis}
\label{analysis}

We performed the spectral analysis using {\textsc{xspec}} (version 12.10.1f) \citep{1996ASPC..101...17A}, and we quote the $1-\sigma$ errors on the best-fit spectral parameters. We consider best-fit models as statistically acceptable when $p_{\rm null}\ge 0.01$.

\subsection{X-ray spectral analysis}
\label{xrayspec}

\begin{table*}
\centering
	\caption{The best-fit parameters from the joint SXT $+$ LX20 spectral fit with the model {{\textsc{const$\times$tbabs$\times$ztbabs$\times$zxipcf$\times$(zpowerlaw+xillver+bbody)}}}.}
	\label{tab:xray_spec}
	\begin{tabular}{cccccccc}
		\hline\hline		
		Model	& Component         &Parameter$^a$    &\multicolumn{5}{c}{ObsID} \\
		& &&9000001006 &9000001048 &9000001118&9000001286&9000001340\\
		\hline	
&	{{\textsc{const}}}&&0.82$^{+ 0.01}_{- 0.01}$ &0.82$^{*}$&0.82$^{*}$&0.82$^{*}$&0.82$^{*}$\\

		\\
		
		&{{\textsc{ztbabs}}}  &$N_{H}$(10$^{22}$ cm$^{-2}$) &0.17$^{+ 0.01}_{- 0.01}$  &0.17$^{*}$ &0.17$^{*}$ &0.17$^{*}$ &0.17$^{*}$\\
		\\\
	    &{{\textsc{zxipcf}}}  &$N_{HW}$(10$^{22}$ cm$^{-2}$) &2.35$^{+ 0.09}_{-0.08}$   &2.35$^{*}$ &2.93$^{+0.18}_{-0.12}$ &2.35$^{*}$ &2.35$^{*}$\\
							  &&$\log{\xi^b}$	&0.46$^{+0.06}_{-0.05}$    &0.46$^{*}$ &0.46$^{*}$ &0.46$^{*}$ &0.46$^{*}$\\
							  &&$CF$			&$0.88^{+0.01}_{-0.01}$    &$0.88^{*}$ &$0.88^{*}$ &$0.88^{*}$ &$0.88^{*}$\\
		\\\
	&	{{\textsc{zpwerlaw}}}  &$\Gamma$    &1.89$^{+ 0.02}_{- 0.02}$	&1.85$^{+ 0.03}_{- 0.03}$	&1.77$^{+ 0.03}_{- 0.03}$	&1.95$^{+ 0.02}_{- 0.02}$	&1.98$^{+ 0.02}_{- 0.02}$\\
		
	&	&$N_{PL}^c$($10^{-2}$)	&4.85$^{+ 0.06}_{- 0.06}$	&4.49$^{+ 0.07}_{- 0.07}$	&2.27$^{+ 0.03}_{- 0.03}$	&5.62$^{+ 0.07}_{- 0.07}$	&6.07$^{+ 0.07}_{-0.07}$\\
		
	&  & $f_{PL}$($2-10\kev$)$^d$ &1.45$^{+0.02}_{-0.02}$	&1.44$^{+0.02}_{-0.02}$	&0.83$^{+0.01}_{-0.01}$	&1.54$^{+0.02}_{-0.02}$	&1.59$^{+0.02}_{-0.02}$\\
		\\
	
	& {{\textsc{xillver}}}	
		
		&$N_{xill}^e$($10^{-4}$) &3.07$^{+0.30}_{-0.30}$&6.44$^{+0.52}_{-0.52}$&3.07$^{*}$&3.07$^{*}$&3.07$^{*}$\\
		
		\\
		
	&	{{\textsc{bbody}}} &$kT_{BB}$(\keV{}) &0.258$^{+0.005}_{-0.005}$ &0.258$^{*}$	&0.258$^{*}$	&0.258$^{*}$	&0.258$^{*}$\\
		
	&	&$N_{BB}^f$(10$^{-3}$) & 1.63$^{+0.24}_{-0.23}$ &1.82$^{+ 0.25}_{- 0.23}$&1.41$^{+0.20}_{-0.18}$&2.08$^{+0.29}_{-0.28}$&2.07$^{+0.30}_{-0.28}$\\
		
	&	 & $f_{BB}$($0.3-2\kev$)$^d$ &1.31$^{+0.18}_{-0.18}$ &1.46$^{+0.18}_{-0.18}$ &1.10$^{+0.14}_{-0.14}$ &1.66$^{+0.22}_{-0.22}$ &1.65$^{+0.22}_{-0.22}$\\
		\\
		\hline
		
	&	$\chi^{2}/dof$ &&&&&&2190.83/2166\\
		
		\hline\hline
	\end{tabular} \\
	$^a$ The notations "f" and "*" indicate fixed and tied parameters, respectively.\\ 
	$^b$ Ionization parameter in unit of ergs~cm~s$^{-1}$.\\	$^c$ Power-law normalization in units of photons~keV$^{-1}$~cm$^{-2}$~s$^{-1}$ at 1 keV\\
	$^d$ Intrinsic X-ray flux in units of $10^{-10}{\rm~ergs~cm^{-2}~s^{-1}}$.\\
	$^e$ Normalization of the {\textsc{xillver}} model.\\
	$^f$ Normalization of the {\textsc{bbody}} model in units of $L_{39}/D_{10}^2$, where $L_{39}$ is the source luminosity in units of $10^{39}$ erg~s$^{-1}$ and $D_{10}$ is the distance to the source in units of 10 kpc\\
\end{table*}

\begin{figure*}
    \centering

  \includegraphics[scale=0.53]{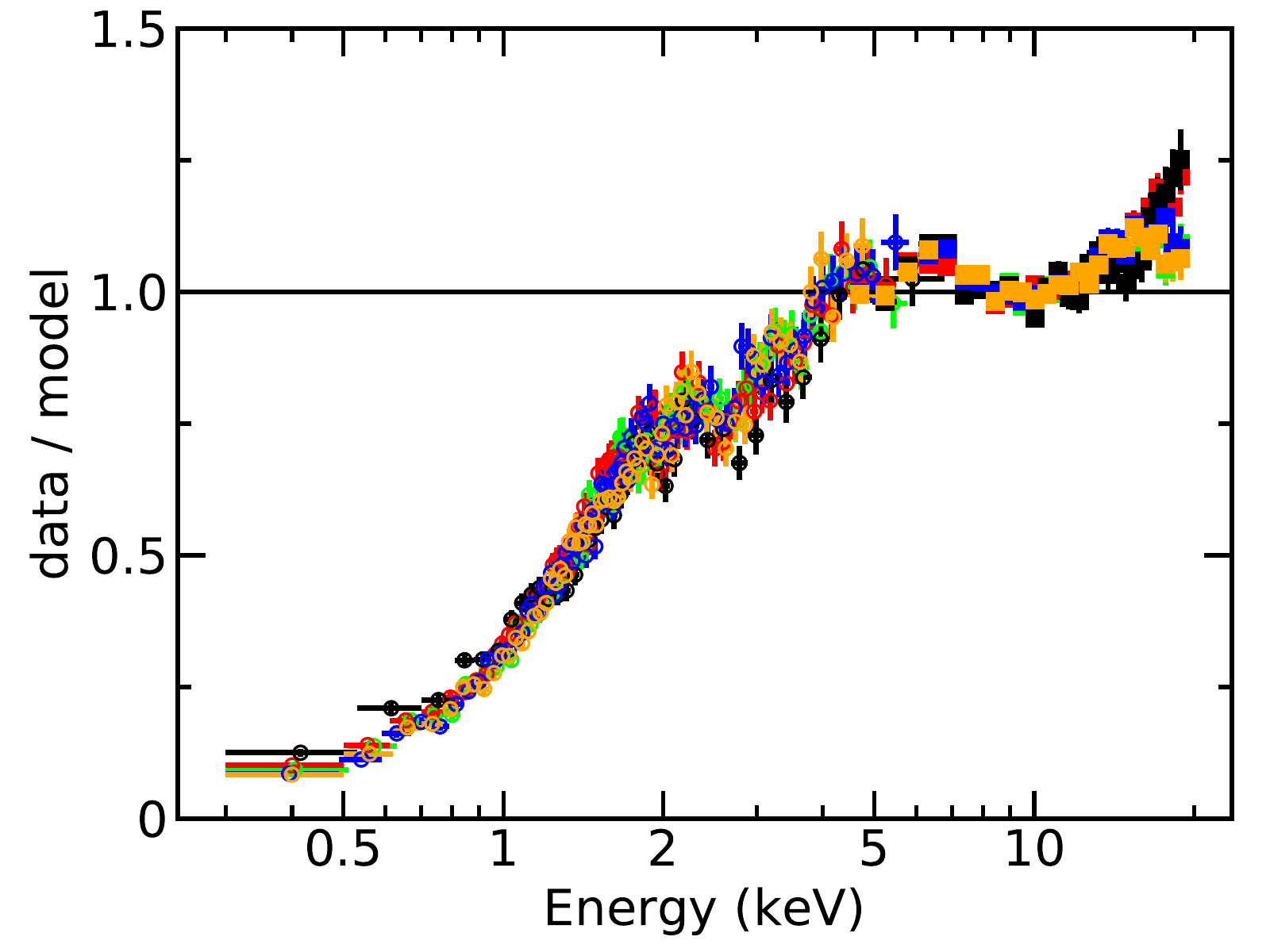}
  \includegraphics[scale=0.53]{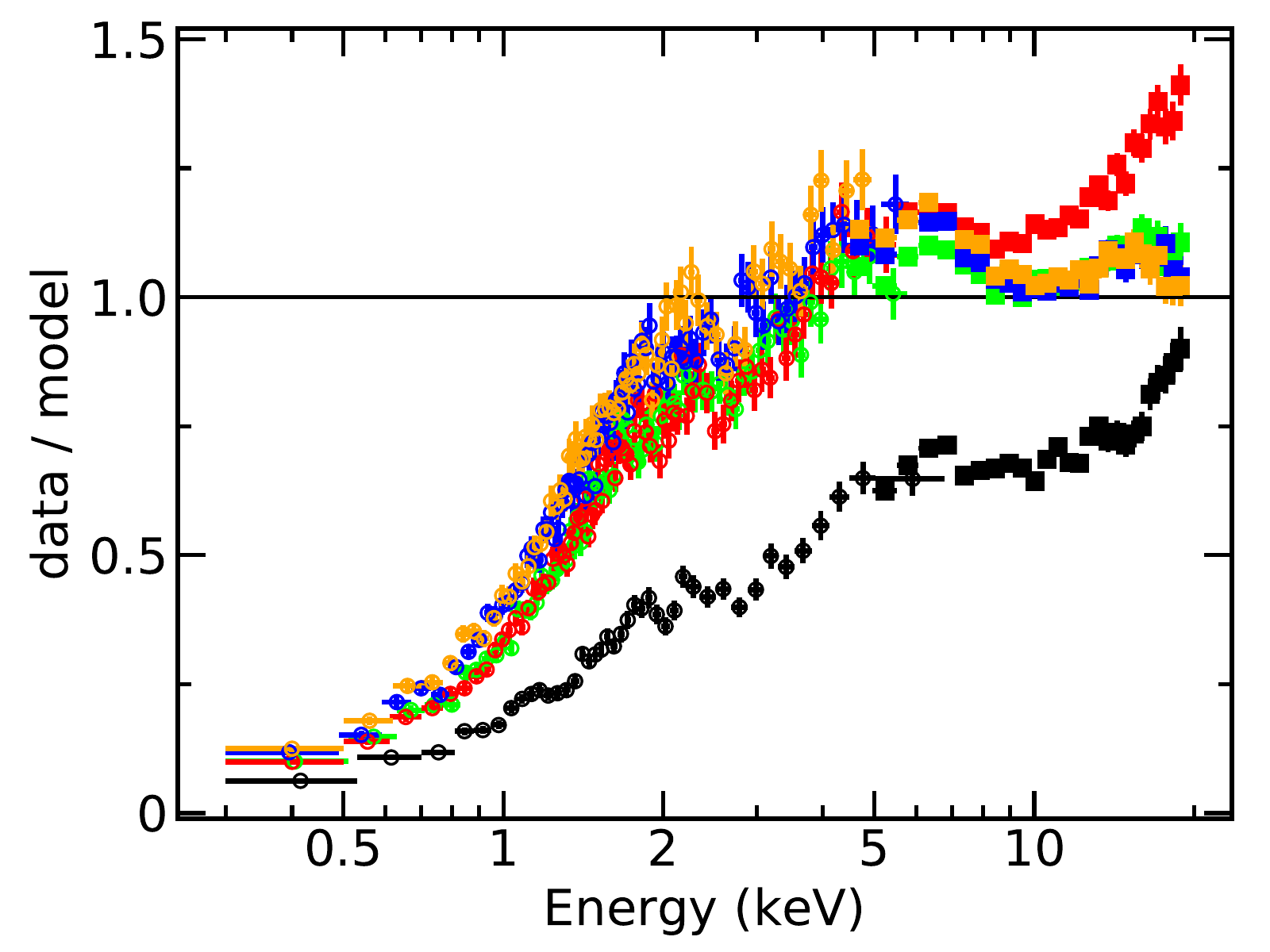}
  \caption{{The SXT ($0.3-7.4\kev$) and LAXPC ($4-20\kev$) spectral data (open circles and filled squares, respectively) from five observations. The spectral data are divided by the  best-fitting {\textsc{powerlaw}} model fitted jointly to the $4-5$ and $8-12\kev$ bands, when the photon-index and normalization are allowed to vary for each observation ({\it left panel}), and when they are tied across the 5 datasets ({\it right panel}). Green, red, black, blue and orange points mark the ObsIDs 9000001006, 9000001048, 9000001118, 9000001286 and 9000001340, respectively (the same colour code and point symbols are used in all relevant figures in this work).}}
    \label{figfirstlook}
\end{figure*}

We used the SXT data in the $0.3-7.4\kev$ band and the LX20 data in the $4-20\kev$ band to ensure good signal-to-noise. {First we fitted the five sets of SXT and LX20 spectral data jointly in the $4-5\kev$ and $8-12\kev$ bands, only. These bands are dominated by the intrinsic continuum emission. We used a simple {\textsc{powerlaw}} model where we varied the photon index and normalization for each dataset. This fit resulted in $\chi^2=214.79$ for 240 degrees of freedom (dof) with $\Gamma \sim 1.7-1.9$ and normalization in the range of $\sim (1.9-5.0)\times10^{-2}$ photons~cm$^{-2}$~s$^{-1}$~keV$^{-1}$ at 1\kev{}. The left panel in Figure~\ref{figfirstlook} shows the ratio of the SXT and LX20 data over the best-fit {\textsc{powerlaw}} model for each of the five observations. The spectra are clearly affected by significant absorption. We can also see the the iron line and the reflection hump at higher energies (which may even be variable). Then we re-fitted the data in the same energy bands, but this time we kept  $\Gamma$ and normalization tied across the five datasets. The right panel in  Fig.~\ref{figfirstlook} shows the data/model ratio for this case.  This plot shows clearly the significant flux variability of the source during the \astrosat{} observations. We detect a significantly smaller flux during ObsID: 9000001118. The spectral slope during this observation is clearly flatter than in the other four observations. The flux (and spectral) variations during the other observations are of smaller amplitude. We discuss below the results from the joint spectral analysis of the five SXT+LX20 spectral datasets. }

%

We note that after the launch of \astrosat{}, the SXT gain has been found to shift slightly. We handled this gain change by shifting the energies in the SXT RMF and ARF with the help of the {\textsc{xspec}} command {\textsc{gain}}. We fixed the gain slope to 1.0 and varied the intercept parameter. For each observation, we tied all the parameters of the spectral models for the LX20 and the SXT spectra.  We also used a constant to account for any difference in the relative normalization of the SXT and LAXPC spectra, and we applied $3\%$ systematic errors in the spectral models to account for any residual calibration uncertainties.

{\it \underline{Model fitting the 2--20 keV band.}} We begin our spectral analysis by fitting the $2-20\kev$ band with an absorbed power-law model ({{\textsc{tbabs$\times$zpowerlaw}}} in {\textsc{xspec}} terminology), with the absorption column density fixed at the Galactic value ($N_H=4.61 \times 10^{20} \cm^{-2}$; \citealt{2005A&A...440..775K}). We used the absorption cross-section and abundances from \cite{1996ApJ...465..487V} and \cite{2009ARA&A..47..481A}, respectively. We varied both the normalization and photon-index of the power-law model for each observation. The model resulted  in a poor fit with $\chi^2= 1750.3$ for $1331$ dof. We added an intrinsic neutral absorption model ({{\textsc{ztbabs}}}) at the source redshift, and we kept the absorption column tied among all five datasets. The fit improved significantly ($\Delta\chi^2 = 212.7$ for one extra dof). { We investigated if the neutral absorbing component is variable, and therefore untied the column density across the five datasets which did not improve the fit significantly ($\Delta\chi^2 = 7.6$ for 4 extra parameters; $p_{\rm null}=0.16$). Therefore, we kept this parameter tied.} 

We then used the reflection model {{\textsc{xillver}}} \citep{xillver1} to account for the iron line and related reflection emission. We tied the photon-index of the reflection component with those of the power-law component for each observation, and we kept the normalization of the {\textsc{xillver}} model tied across 5 datasets. {We fixed the iron abundance at $A_{Fe}=1$, inclination angle at $\theta=60^{0}$, ionization parameter at $\log{\xi}=0$, cut-off energy at $E_{cut}=186\kev$ \citep{2014ApJ...788...61B}, and reflection fraction at $R_f=-1$}. This resulted in a better fit with $\chi^2/dof = 1457.2/1329$. We investigated if the {{\textsc{xillver}}} component is variable, and we untied its normalization. We found that the normalization was almost constant in 4 observations and increased by a factor of $\sim2$  in one of them (ObsID: 9000001048). Therefore, we tied the normalization in 4 observations and allow it to vary freely in the other one, which resulted in an improved fit with $\chi^2/dof=1446.4/1328$. {We note that this variation cannot be in response to the observed X-ray continuum variation between the first and second observations, since the power-law component did not vary. If we assume that the X-ray reflection originates in the obscuring torus which may be located $\sim 10^5 R_g$ away from the central source (at least), then for a $\sim 10^8$ M$_{\odot}$ black-hole, we expect to see correlated variations on a timescale of more than a year, which is much longer than this campaign.} The best-fit results for the fit to the 2--20 keV band are listed in Table~\ref{tab:xray_spec}. 

{\it \underline{Model fitting the full X-ray band.}} Then we noticed the 0.3--2\kev{} band in all SXT datasets in order to study the spectrum in the soft X--ray band. First, we fitted the 0.3--20\kev{} band with the best-fit model that we derived from the fits to the 2--20\kev{} band. This resulted in a poor fit ($\chi^2/dof=4138.8/2176$). Inspection of the best-fit residuals indicated the presence of a soft X-ray excess emission. We therefore added a black-body  component {{\textsc{bbody}}} to the model, and we repeated the fit, keeping the power-law photon index fixed at the best fit values we derived when fitting the hard band spectra. Initially we tied the {{\textsc{bbody}}} temperature ($kT_{BB}$) and normalization across the 5 datasets. This resulted in an improved but still statistically poor fit with $\chi^2/dof=2663.3/2174$. The fit improved to $\chi^2/dof=2530.7/2170$ when we allow the {{\textsc{bbody}}} normalization to vary freely. We also investigated  if $kT_{BB}$ varies as well. We found that the black-body temperature is similar, within errors, for all but one observation (ObsID: 9000001118), where the best-fit the temperature is different. We untied $kT_{BB}$ for this observation, and this further improved the fit ($\Delta\chi^2=17.1$ for 1 dof). 

\begin{figure*}
	\centering
	\includegraphics[scale=0.53]{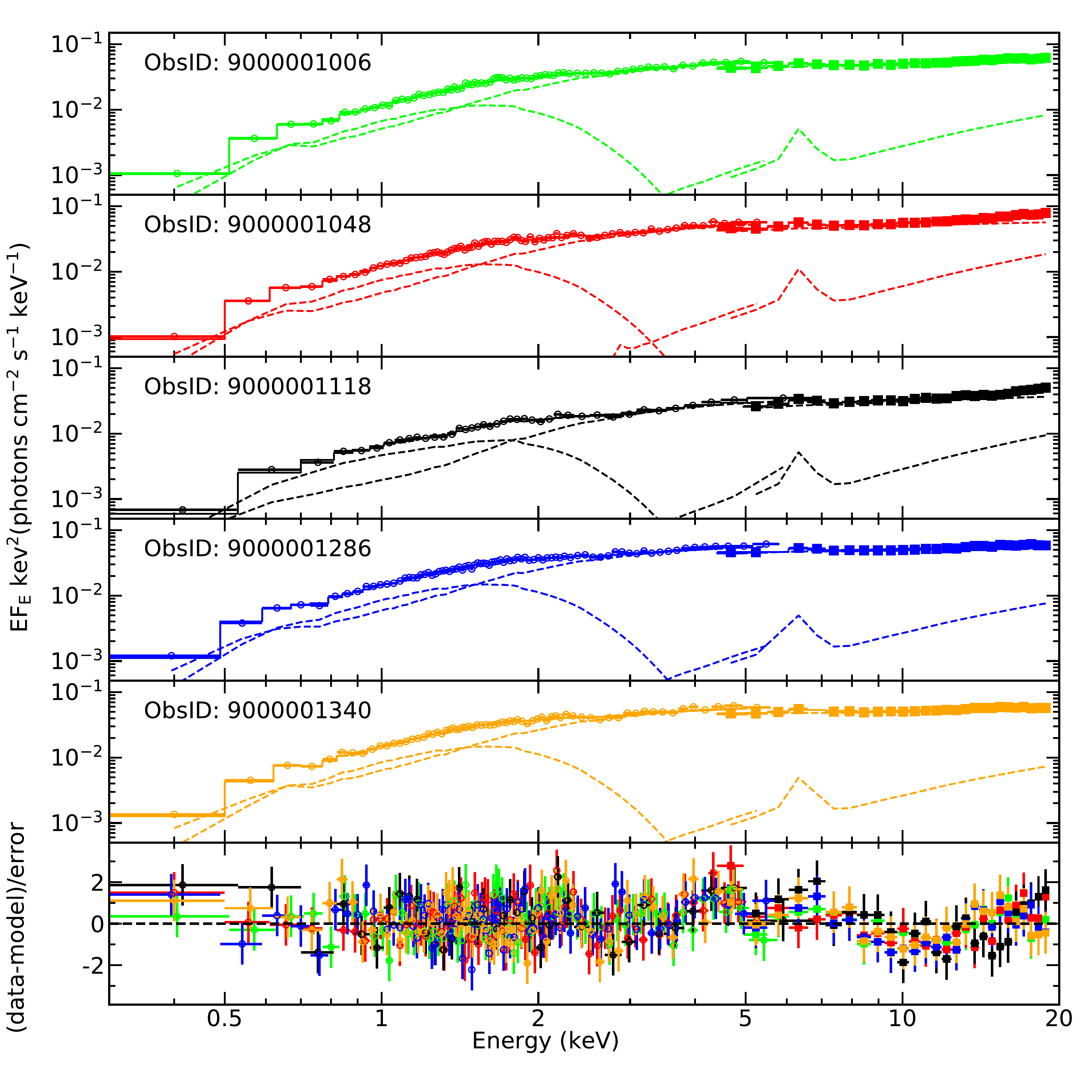}
	\caption{The SXT ($0.3-7.4\kev$) and LAXPC ($4-20\kev$) spectra of the 5 observations. Top 5 panels show the data and the best-fit models for individual observation (the observation ids are shown) and the bottom panel shows $\chi$ residuals in terms of (data$-$model)/error. Solid and dashed lines in each panel show the overall best-fit models, and the individual spectral components, respectively (see \S~\ref{xrayspec} for details).}
	
\label{fig2}
\end{figure*}

The best-fit residuals indicated the presence of further absorption, due to ``warm" material this time. We therefore added a warm absorption component, {\textsc{zxipcf}}, to the model. We tied all the parameters of the absorption component across 5 datasets, and the resulting fit improved significantly, to $\chi^2/dof=2207.21/2166$, which is a statistically acceptable solution ($p_{\rm null}=0.26.$). { We investigated if the parameters of the warm absorbing component vary across the 5 datasets. First we checked if the ionization parameter, log($\xi$), is variable across the 5 datasets. We kept the column density, $N_{HW}$, and the covering factor ($CF$) tied, and let the ionization parameter to vary during the fit. This did not improve the quality of the fit significantly ($\Delta\chi=1.69$ for 4 extra dof). Then, we kept $N_{HW}$ and log($\xi$) tied and we let $CF$ to vary during the fit. This did not improve the fit either ($\Delta\chi = 5.45$ for 4 parameters). Then, we kept $CF$ and log$(\xi)$ constant, and let $N_{HW}$ to be variable. We noticed that the column density is similar for 4 observations but different for ObsID: 9000001118. We therefore tied the column density for the 4 observations, and we repeated the fit. The resulting $\chi^2$ of 2190.8/2165 dof indicates that the quality of the fit improved significantly in this case (we find an F-statistic of 16.2 and $p_{\rm null}\sim 6\times 10^{-5}$ when we apply the F-test). At the same time, we also found that the black-body temperature  for ObsID: 9000001118 is now consistent with those for the four other observations. Hence, we tied $kT_{BB}$ among all five observations and we repeated the fit. This resulted in a  $\chi^2$ of 2190.8 for 2166 dof, which indicates a very good fit ($p_{\rm null}=0.35)$.} We accept this as our final fit of the broad-band X--ray spectrum of the source during the  five \astrosat\ observations. The best-fit parameters are listed in Table~\ref{tab:xray_spec}, together with the intrinsic power-law flux in the 2--10 keV band, and the black-body flux in the 0.3--2 keV band, which we consider as indicative of the soft-excess flux. Figure~\ref{fig2} shows the spectra, the overall best-fit model and the individual components. The bottom panel shows the model residuals in terms of sigmas with error bars of size one. 

Our results show that the power-law photon index varied from $\sim1.8$ to $\sim2$ during the \astrosat\ monitoring of the source. The power-law and the soft X-ray excess fluxes are also variable by a factor of $\sim2$ and $\sim1.5$ respectively, though the errors on the soft X-ray excess fluxes are relatively larger (see Table~\ref{tab:xray_spec}). The  fractional root mean square (rms) variability amplitude \citep{2003MNRAS.345.1271V} of the X--ray power-law and the soft-excess flux, using the $f_{PL}$ and the $f_{BB}$ values listed in Table~\ref{tab:xray_spec}, are 22.46$\pm 0.60$\% and 9.87$\pm 5.93$ \%, respectively. Interestingly, the X--ray continuum variability amplitude is slightly {\it smaller} than the NUV rms, which is 26.20$\pm 2.24$ \% (using the intrinsic AGN count rates listed in Table~\ref{tab:nuvres}), although the difference is not statistically significant. Nevertheless, this is rather unusual for AGN.  

 \begin{figure}
 	\centering
 \includegraphics[scale=0.42]{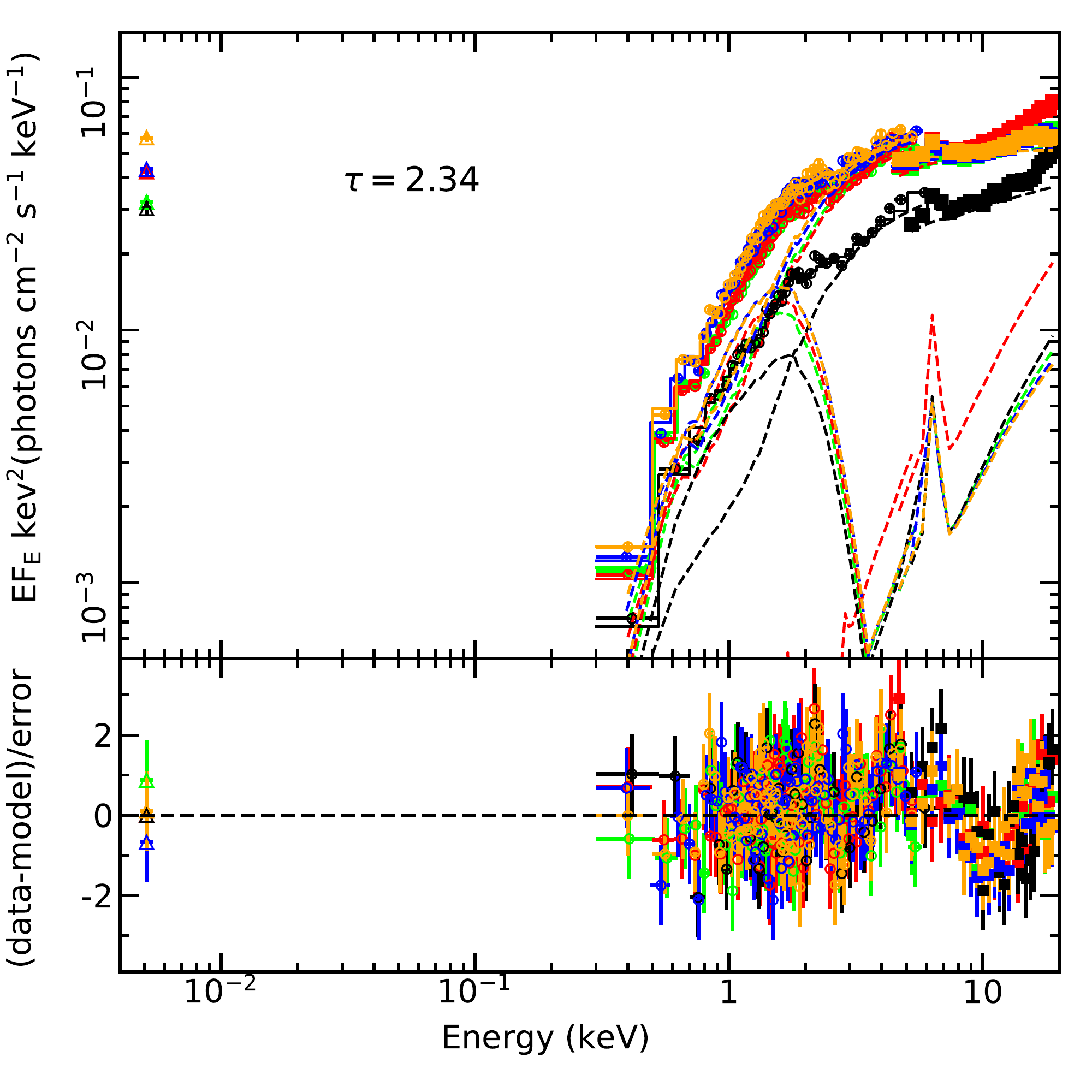}
 \caption{The NUV/X-ray SEDs for all observations, fitted with the {{\textsc{const$\times$phabs$\times$zphabs$\times$zxipcf$\times$(thcomp$*$optxagnf+ bbody+xillver)}}} model when the optical depth is fixed and the corona temperature varies. 
 The bottom panels show $\chi$ residuals in terms of (data$-$model)/error). The NUV data are shown in open triangles.}
 \label{fig6}
 \end{figure}

\subsection{Broadband UV/X-ray spectral analysis}
\label{broadbandspec}

We investigated the UV/X-ray broadband spectral variability of IC~4329A by constructing and fitting the broadband SEDs using the intrinsic UV  and the X-ray spectral data acquired simultaneously with \astrosat{}. We converted the intrinsic UV flux into {\textsc{xspec}} compatible spectral files for each observation (Table~\ref{tab:nuvres}) using the {\textsc{ftflx2xsp}}{\footnote{\url{https://heasarc.gsfc.nasa.gov/lheasoft/ftools/headas/ftflx2xsp.html}}} tool. 
 
We added an accretion disk component to the model ({\textsc{optxagnf}};  \cite{10.1111/j.1365-2966.2011.19779.x}) in order to fit the UV/X-ray data. The main parameters of the {\textsc{optxagnf}} are$\colon$ the black-hole mass ($M_{BH}$), the comoving/proper distance ($D$; in\mpc{}), the Eddington ratio ($\dot{m}=\log(L/L_{Edd})$), the dimensionless black-hole spin ($a$), and the inner disk radius ($R_{cor}$, in $R_g = GM_{BH}/c^2$). We forced the {\textsc{optxagnf}} model to produce the accretion disk spectrum only, by setting $R_{cor}$ to be negative. 

\cite{Dewangan_et_al_2021} studied in detail the optical/UV spectrum of IC~4329A using the same \astrosat{} observations that we also consider in this work. They corrected the optical/UV data for the intrinsic extinction using multiple extinction curves, and they modeled the intrinsic optical/UV data with the {\textsc{optxagnf}} disk model both for a maximally spinning Kerr and a Schwarzschild black-hole. They showed that the disk must be truncated in both cases. {Here, we assume their best-fit results derived by fitting the Kerr disk model to the intrinsic extinction-corrected data using the extinction curve of \cite{2004MNRAS.348L..54C} with a color excess of $E(B-V)=1.0$. We therefore fixed the inner disk radius at $ R_{cor} = 80R_g$, and we fitted the data assuming $M_{BH} = 2\times10^{8}M_{\odot}, D = 69.61\mpc,$ and $a = 0.998$, like \cite{Dewangan_et_al_2021}. We note that the the best-fit inner radius is the same irrespective of whether we assume a non-rotating or a maximally rotating black-hole (what changes in these cases is the best-fit accretion rate value). This is because the differences between the spectrum of an accretion disk around a spinning Kerr and a Schwarzschild black-hole differ at wavelengths shorter than $\sim 1000$\angstrom{}, where the intrinsic emission of IC~4329A is absorbed.}

Furthermore, we replaced the {\textsc{zpowerlaw}} component with {\textsc{thcomp}}, which  is a convolution model that can be used to Comptonize any seed photon spectrum \citep{2020MNRAS.492.5234Z}. The main parameters of this component are$\colon$ (1) the X-ray power-law photon index $\Gamma$ or the Thomson optical depth $\tau$, (2) the electron temperature $kT_e$, and (3) the scattering fraction $f_{sc}$ (where, $0 \leq f_{sc}\leq 1$; $f_{sc} =1$ implies that all seed photons will be Comptonized). We convolved the disk spectrum produced with the {\textsc{optxagnf}} with the {\textsc{thcomp}} model. We also replaced the {\textsc{tbabs}} and {\textsc{ztbabs}} components by {\textsc{phabs}} and {\textsc{zphabs}}, respectively, as the former components also account for UV absorption. Thus, the final model is {\textsc{const$\times$phabs$\times$zphabs$\times$zxipcf$\times$(thcomp$*$optxagnf +bbody+xillver)}}. We used this model to fit all five NUV and the broad-band X-ray spectra of the source. We fixed the parameters of {\textsc{bbody}}, {\textsc{xillver}}, {\textsc{zxipcf}}, and {\textsc{zphabs}} at the best-fit values that we derived from the X-ray spectral fits, and allowed the accretion rate and the parameters of the Comptonization component to be variable during the fit. Our objective is to investigate if this model, with the more physical model components, can fit the NUV/X-ray spectra of the source. 

   \begin{table*}
   	\centering
   	\caption{Results of the joint UV/X-ray broadband spectral fitting. {We note that the logarithm of the mass accretion rate is in terms of the Eddington accretion rate, i.e. $\log(\dot{m})=\log(\dot{M}/\dot{M}_{\rm Edd})$, since this is equal to $\log(L/L_{Edd})$.}}
   	\label{tab:thcomp}
   	\begin{tabular}{ccccccc} \hline\hline
   		ObsID &\multicolumn{3}{c}{$kT_e=50\kev$} &  \multicolumn{3}{c}{$\tau=2.34$}  \\
   		& $\log{\dot{m}}$  & $\tau$ & $f_{sc}$ & $\log{\dot{m}}$  & $kT_e$ & $f_{sc}$   \\ \hline
   	
   		1006  	& $-1.48^{+0.04}_{-0.04}$ &  $1.80^{+0.01}_{-0.01}$	&  $0.79^{+0.08}_{-0.07}$ & $-1.47^{+0.04}_{-0.03}$ &  $35.33^{+0.25}_{-0.24}$	& $0.71^{+0.05}_{-0.06}$\\

   		1048 &  $-1.15^{+0.03}_{-0.03}$ & $1.90^{+0.01}_{-0.01}$ & $0.26^{+0.02}_{-0.02}$ &  $-1.15^{+0.02}_{-0.02}$ & $37.84^{+0.36}_{-0.34}$	& $0.25^{+0.02}_{-0.01}$	\\

   		1118 & $-1.19^{+0.03}_{-0.03}$ & $2.06^{+0.02}_{-0.02}$& $0.097^{+0.009}_{-0.009}$ & $-1.19^{+0.03}_{-0.03}$ & $41.98^{+0.60}_{-0.54}$	&  $0.095^{+0.009}_{-0.008}$	 \\

   		1286 &  $-1.30^{+0.04}_{-0.04}$ &  $1.71^{+0.01}_{-0.01}$ & $0.79^{*}$ & $-1.30^{+0.04}_{-0.02}$ & $33.25^{+0.17}_{-0.22}$	&  $0.71^{*}$ \\

   		1340 & $-1.19^{+0.03}_{-0.04}$ & $1.67^{+0.01}_{-0.01}$ & $0.79^{*}$  &  $-1.19^{+0.04}_{-0.04}$ & $32.05^{+0.21}_{-0.22}$   & $0.71^{*}$	\\ \hline
   		
   	\end{tabular}
   \end{table*}

However, it is difficult to determine both the electron temperature and the Thomson optical depth of the corona due to the lack of X-ray data at high energies, where the energy cut-off could be observed. For that reason, we fitted the data twice. In the first case, we fixed $kT_e$ at $50\kev$ \citep{2014ApJ...788...61B} assuming a spherical geometry of the corona, we tied $f_{sc}$,
and we let $\tau$ and the accretion rate be variable during the fit. The best-fitting model resulted in $\chi^2/dof = 2413.18/2179$. We tested the possibility of variable $f_{sc}$, and we found that the fit improves significantly ($\Delta\chi^2=114.15$, for an additional parameter) when we allow $f_{sc}$ to vary freely in one observation (ObsID: 9000001118). The fit improved further ($\Delta{\chi^2}= 77.27$ for an extra parameter), if we allow $f_{sc}$ to vary for a second observation (ObsID: 9000001048). We did not notice any further improvement if we allow $f_{sc}$ to vary in the other observations. The final model resulted in a $\chi^2/dof = 2221.76$ for $2177$ dof. The best-fit results are listed in Table~\ref{tab:thcomp}.
Next, we fixed the optical depth at $\tau = 2.34$ \citep{2014ApJ...788...61B} and we allow the accretion rate and electron temperature, $kT_e$, to vary during the fit. The model again resulted in a good fit with $\chi^2/dof = 2221.80/2177$ dof. As with the fit above, the model fits the data best when we allow $f_{sc}$ to vary freely for the first and second observation. The variable $kT_e$ variant of the model fits the data as well as when we let $\tau$ to vary. The best-fit parameters are listed in Table~\ref{tab:thcomp}. In order to demonstrate the quality of the model fits in these cases, we show the best-fit UV/X-ray SED when $\tau$ was kept fixed in Figure~\ref{fig6}. 

\subsection{The soft X-ray excess and the disk emission} 
\label{sec:se_ad}

We also investigated whether the soft-excess component can be explained as thermal Comptonization of the disk photons in a 'warm' corona. For this, we revisited the joint X-ray spectral analysis. We removed the {\textsc{bbody}} model and used the {\textsc{nthcomp}} model instead \citep{nthcomp1,nthcomp2} to fit the soft X-ray excess emission. Thus, our model now becomes {\textsc{const$\times$tbabs$\times$ztbabs$\times$zxipcf$\times$(zpowerlaw+ xillver+nthcomp)}}. The main parameters of the {\textsc{nthcomp}} model are$\colon$ photon-index ($\Gamma_{warm}$) of the Comptonized spectrum, electron temperature ($kT_{warm}$), and seed photon temperature ($kT_{seed}$). We assumed black-body seed photons (int\_type $=0$) with the surface temperature profile of the disk given in equation~\ref{eqn4}, and we calculated the seed photon temperature at $R_{in} = 80R_g$ for each observation, using the mass accretion rate given in Table~\ref{tab:thcomp}, and a black-hole mass of $2\times10^8 M_{\odot}$. {The seed photon temperatures ($kT_{seed}$) are, $0.85, 1.03, 1.01, 0.95,$ and $1.01\ev$ in the sequence of the observation ids.}

 We kept $\Gamma_{warm}$, $kT_{warm}$, and normalization  of the {\textsc{nthcomp}} model ($N_{nthcomp}$) tied during the model fit. This resulted in $\chi^2/dof = 2289.46/2169$. The quality of the fit improved considerably when we allow the normalization to vary freely during the fit, resulting in $\chi^2/dof = 2189.45/2165$.
 The quality of this fit is almost identical to the quality of the model fit when we used the {\textsc{bbody}} model to account for the soft excess. The best-fit values of the {\textsc{nthcomp}} parameters are$\colon$ $\Gamma_{warm}\leq 2.54$ ($3\sigma$ upper limit), $kT_{warm} =0.263^{+0.005}_{-0.005}\kev$, and $N_{nthcomp} = 6.12^{+0.95}_{-0.84} \times 10^{-2}$, $6.81^{+0.98}_{-0.87} \times 10^{-2}$, $5.36^{+0.73}_{-0.67} \times 10^{-2}$, $7.78^{+1.15}_{-1.03} \times 10^{-2}$, and $7.75^{+1.10}_{-1.05} \times 10^{-2}$ ($N_{nthcomp}$ = unity at 1\kev{} for a norm of 1) in the sequence of the observation ids. Other parameters remain unchanged. 
 
{Lastly, we also tried to fit simultaneously the NUV, soft X-ray excess, and power-law components of the 5 datasets using the \textsc{optxagnf} model. In {\textsc{xspec}} terminology, the model is now: {\textsc{const$\times$phabs$\times$zphabs$\times$zxipcf$\times$(optxagnf+xillver)}}. We fixed the photon-index ($\Gamma$) of the {\textsc{optxagnf}} model at the best-fit values listed in Table~\ref{tab:xray_spec}. 
We allow the temperature and optical depth of the warm corona, the power-law fraction ($f_{pl}$), and the mass accretion rate for each observation to vary freely in the fit, in order to take into account the variations we saw in the \textsc{BBODY} component, in the power-law and the NUV flux (see \S~\ref{xrayspec} and \ref{uvitanalysis}). We tied the inner disk radius ($R_{cor}$) across the 5 datasets, and we also allow it to vary when we fit the data. The best-fit results show that the power-law fraction is in the range  $0.48-0.69$, the temperature of the warm corona is of the order of $\sim 0.24$ keV, while the optical depth varies between $\sim 20$ and $40$. We found an inner disk radius of $\geq 195 R_g$. The best-fit model resulted in $\chi^2/dof = 2253.16/2165$. This is a good fit ({\it $p_{null}=0.09$}), although it is not as good as the fit to the combined NUV/X-ray spectra we presented in \S~\ref{broadbandspec}. The best-fit inner disk radius is larger than the value we assumed in that Section (which was based on the modelling of the UV/optical SED of the source of \citealt{Dewangan_et_al_2021}). But this could be due to the fact that we may be reaching the accuracy of the models themselves. The important thing is that general picture is the same: a truncated disc, and a warm corona, which vary in accretion rate and (perhaps) in optical depth, can explain the observations. }


\section{Results \& Discussion}
\label{discussion}

We analyzed five simultaneous UV/X-ray observations of IC~4329A  performed with \astrosat{} during February -- June 2017, and we investigated the broadband X-ray spectral variability and its connection with the NUV emission from the AGN.  The excellent spatial resolution of UVIT  ($\sim 1-1.5\arcsec$) allowed us to reliably separate the AGN from the host galaxy emission by fitting the radial profile of IC~4329A for each observation separately. We corrected the AGN emission for Galactic and internal reddening using most suitable extinction law derived for IC~4329A.  We also corrected for the contribution of BLR/NLR and derived the intrinsic continuum emission from the AGN in the NUV band (see, section~\ref{uvitanalysis} and Table~\ref{tab:nuvres}). Light curves generated from simple aperture photometry as well as from the radial profile analysis show similar variability patterns, and demonstrate that the disk emission  from IC~4329A varied genuinely by a factor of $\sim 2$ during the five month period of monitoring observations. 

We found that X-ray emission from IC~4329A consists of a continuum, power-law like component with a variable slope ($\Gamma \sim 1.8-2$), a soft X-ray excess component described by a simple black-body ($kT_{BB}\sim 0.26\kev$), and X-ray reflection. The measured temperature of the black-body component is  consistent with the values generally observed for type--1 AGN \citep{2018MNRAS.480.1247K}. We also found that the nuclear emission suffers from both neutral and warm absorption, thus confirming earlier studies (\citealt{2004ApJ...608..157M, 2016MNRAS.458.4198M, 2018A&A...619A..20M}). We converted the column density of the neutral absorbing component ($\rm N_H=1.7\times10^{21}\cm^{-2}$) to optical extinction $A_V=0.77$ using the relation $\rm N_H$ = 2.21$\times10^{21}A_V $ \citep{2009MNRAS.400.2050G}. For R$_V$ = 3.1, this corresponds to a color excess of $E(B-V) = 0.25$. This is  smaller than the color excess of $E(B-V)=1.0\pm0.1$ reported by \citealt{2018A&A...619A..20M} who also pointed  that the neutral gas alone is not sufficient to produce all the observed UV/optical extinction in the source, and dust can also be associated to the low ionized absorbing components. This discrepancy could also be explained if the dust to gas ratio in IC~4329A is larger than that in our Galaxy. 

\begin{figure}
    \centering
    \includegraphics[scale=0.53]{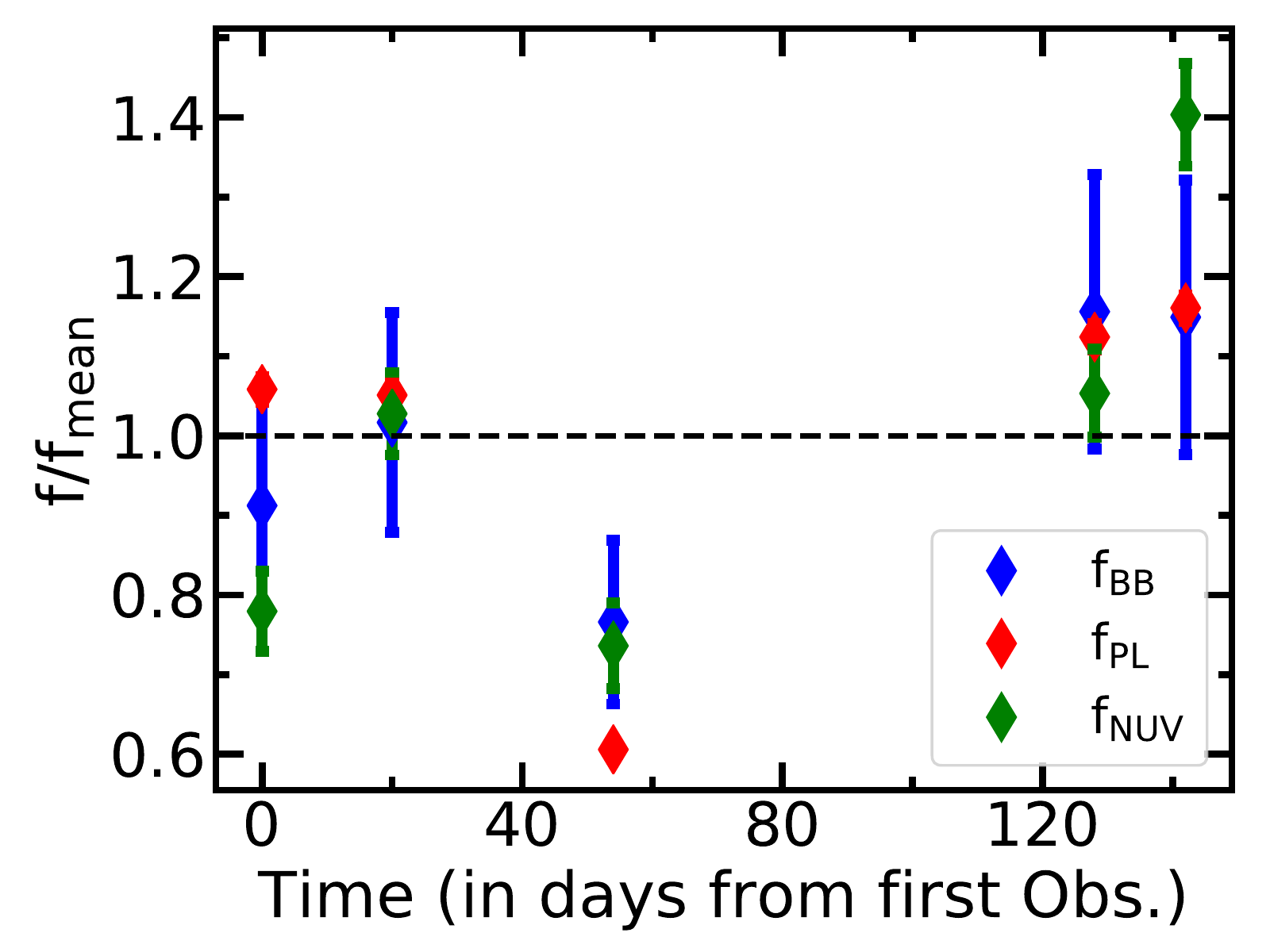}
    \caption{The $2-10$ keV power-law flux (red), the soft X-ray excess flux in the $0.3-2\kev$ band (blue), and the NUV band (green) light curves, normalized to the mean flux of the respective bands.}
    \label{figintrinsicflux}
\end{figure}

\subsection{The variable disk emission}

The NUV flux of IC~4329A is highly variable. We found that the variability amplitude of the disk emission is actually higher than the amplitude of the X-ray variations ($F_{\rm var,NUV}/F_{\rm var,X}\sim 1.2$). { Figure~\ref{figintrinsicflux} shows the intrinsic flux in the NUV band  ($f_{NUV}$), the 0.3--2\kev{} band flux of the black-body component ($f_{BB}$), and the 2--10\kev{} band power-law flux ($f_{PL}$) as a function of time (fluxes are those listed in Table~\ref{tab:nuvres} and \ref{tab:xray_spec}). Each light-curve is normalized to its mean to facilitate the comparison of their amplitude. Firstly, it is clear that the AGN is significantly variable in these bands, and the variations are highly correlated. This is not the result of any  model fitting degeneracy. The NUV flux is determined from the radial-profile analysis of the source in the NUV images, the $f_{PL}$ is determined by the best-fit power-law model in the 2--10 keV band, while $f_{BB}$ is calculated using the best-fit black-body component  while the power-law component is kept fixed. Secondly, the power-law flux is variable by a factor of $\sim 20$\% between the observations, except the third observation where we observe a large amplitude flux decrease. The NUV light curve on the other hand appears to be even more variable than the X--ray continuum flux from one to the next observation. This explains the larger variability amplitude in the NUV band (see \S \ref{xrayspec}). This result is not statistically significant (the variability amplitudes are consistent within the errors), and it is based on just five observations spread over a five-month period. Nevertheless, it is very interesting as it is opposite to what we observe in other nearby AGN. }

Recent results from intensive, multi-wavelength monitoring of a few bright AGN over a period of a few months have shown that the variability amplitude of the {\it Swift} W1 band (which has a central wavelength similar to the  \astrosat{} NUV filter) is always smaller than the X--ray variability amplitude. For example, the $F_{\rm var,NUV}/F_{\rm var,X}$ ratio is of the order of 0.35 in NGC 5548 \citep{2016ApJ...821...56F}, 0.25 in NGC 4593 \citep{2018MNRAS.480.2881M}, 0.4 in NGC~2617 \citep{2018ApJ...854..107F}, 0.15 in NGC~4151 and 0.8 in Mrk~509 \citep{2019ApJ...870..123E}. Except {for} NGC~5548, the $F_{\rm rms,NUV}$ amplitudes may not be corrected for the host galaxy contribution but, at these wavelengths, the host galaxy may not contribute significantly. The large NUV variability amplitude in IC~4329A is even more remarkable, if one considers that this is a high mass AGN. 

Since $F_{\rm var,NUV} \ge F_{\rm var,X}$ in IC~4329A, it is not possible to explain the variable disc emission as the result of X-ray thermal reverberation. Most of the observed NUV variations must be due to intrinsic physical processes in the disk. The dynamical ($t_{dyn}$), thermal ($t_{th}$), and viscous ($t_{vis}$) timescales are given by, \citep{2006ASPC..360..265C}
\begin{equation}
t_{dyn} = \bigg(\frac{r^3}{GM_{BH}}\bigg)^{1/2} \approx 500\bigg(\frac{M_{BH}}{10^8M_{\odot}}\bigg)\bigg(\frac{r}{R_g}\bigg)^{3/2},
\end{equation}

\begin{equation}
t_{th} = \frac{1}{\alpha}t_{dyn},
\end{equation}

and 

\begin{equation}
t_{vis} \approx \frac{1}{\alpha}\bigg(\frac{r}{h}\bigg)^2t_{dyn},
\end{equation}
\noindent
where, $\alpha$ is the viscosity parameter, $r$ is radial distance and $h$ is height of the disk (both in units of $R_g$). We first estimated h/r = $c_s/v_{\phi} \sim 2.9\times 10^{-4}$ (where, $c_s = \sqrt{kT/m_p}$ is the sound speed and $v_{\phi} = \sqrt{GM_{BH}/r}$ is the Keplarian velocity) for a disk temperature of $\sim 11300{\rm~K}$ (calculated below) and a black-hole mass of $M_{BH} = 2\times10^{8}M_{\odot}$. Then, assuming $\alpha=0.1$, we calculated the dynamical, thermal, and viscous timescales of the disk  at $R_{in}=80 R_g$ to be $t_{dyn}\sim8.3$ days, $t_{th}\sim83$ days, and $t_{vis}\sim 2.7\times10^{6}$ years. We find variations by a factor of $\sim 2$ on timescales as short as $\sim 80$ days (see Figure~\ref{figintrinsicflux}). Therefore, they could be due to some disk variation that develops on the thermal timescale. If the warm corona does exist on top of the inner disk (at radii smaller than $\sim 80 R_g$) then perhaps a disk instability at this transition radius, which develops on the thermal timescale, may explain the observed variability. If that is the case, we would expect to see similar amplitude UV variations in other sources, where a warm corona above the disk has been suggested. Perhaps this instability operates over a narrow annulus around the transition radius, and the resulting emission dominates in a narrow spectral range, which in IC~4329A just happens to be within the NUV filter. 
If a thermal instability develops at the transition radius, we would expect the maximum amplitude variations to appear at the wavelength where the maximum flux from this region is emitted. We investigated if the emission from an annulus around the transition radius at $80 R_g$ could contribute to the NUV in IC4329A, by assuming the accretion disk emits as a multi-color black-body. The surface temperature radial profile ($T_s(r)$) of the disk is given by \citep{1999agnc.book.....K}:

\begin{equation}
\label{eqn4}
    T_s (r) = \bigg(\frac{3G\dot{M}M}{8\pi{\sigma}r^3}R_{R}(r)\bigg)^{1/4},
\end{equation}

where, $M$ and $\dot{M}$ are the black-hole mass and the accretion rate in physical units, and  $R_R(r)$ is the general relativistic reduction factor as a function of the radial distance, $r$.
We calculated $\dot{M}$ by the equation: $\dot{M}=\bar{\dot{m}}\dot{M}_{\rm Edd}$, where $\bar{\dot{m}}$ is the mean of the dimensionless accretion rates listed in Table~\ref{tab:thcomp} (which is equal to $\sim 0.06$), and $\dot{M}_{Edd} = L_{Edd}/\eta c^2$, where, $\eta$ is the accretion efficiency and $L_{Edd}$ is the Eddington luminosity. {We assumed $\eta = 0.321$, for a spin parameter of 0.998}. The Eddington luminosity for a mass of $M_{BH} = 1-2\times10^8M_{\odot}$ is $\sim 1.3-2.5\times10^{46}\ergs{}\s^{-1}$. Further, the general relativistic reduction factor is $\sim 0.8$ at $r = 80 R_g$ (see, Figure~7.3 of \citealt{1999agnc.book.....K}). {Using these values, we found that the surface disk temperature at the transition radius is expected to be $\sim 11300-13500{\rm~K}$. Using the Wien's displacement law ($\lambda_{peak}{T_s} = b$, where $b = 2.898\times10^{-3} m.K$ and $\lambda_{peak}$ is the peak wavelength of the black-body spectrum), we estimated that $\lambda_{peak}\sim 2150-2550$\angstrom{}}, which after accounting for the source redshift is  well within 
the bandpass of the NUV/N245M filter \citep{2020AJ....159..158T}. This simple calculation above shows that, if a thermal instability develops at the transition region in IC~4329A, then its variability { timescale} is compatible with what we observe, and the maximum variability amplitude should appear in the NUV band. Perhaps, this can also explain the fact that, while the NUV flux is highly variable, the visible flux 
is almost constant \citep{2018A&A...619A..20M} . 

 
\subsection{The X--ray spectral variability} 

\begin{figure}
 	\centering
 	\includegraphics[scale=0.53]{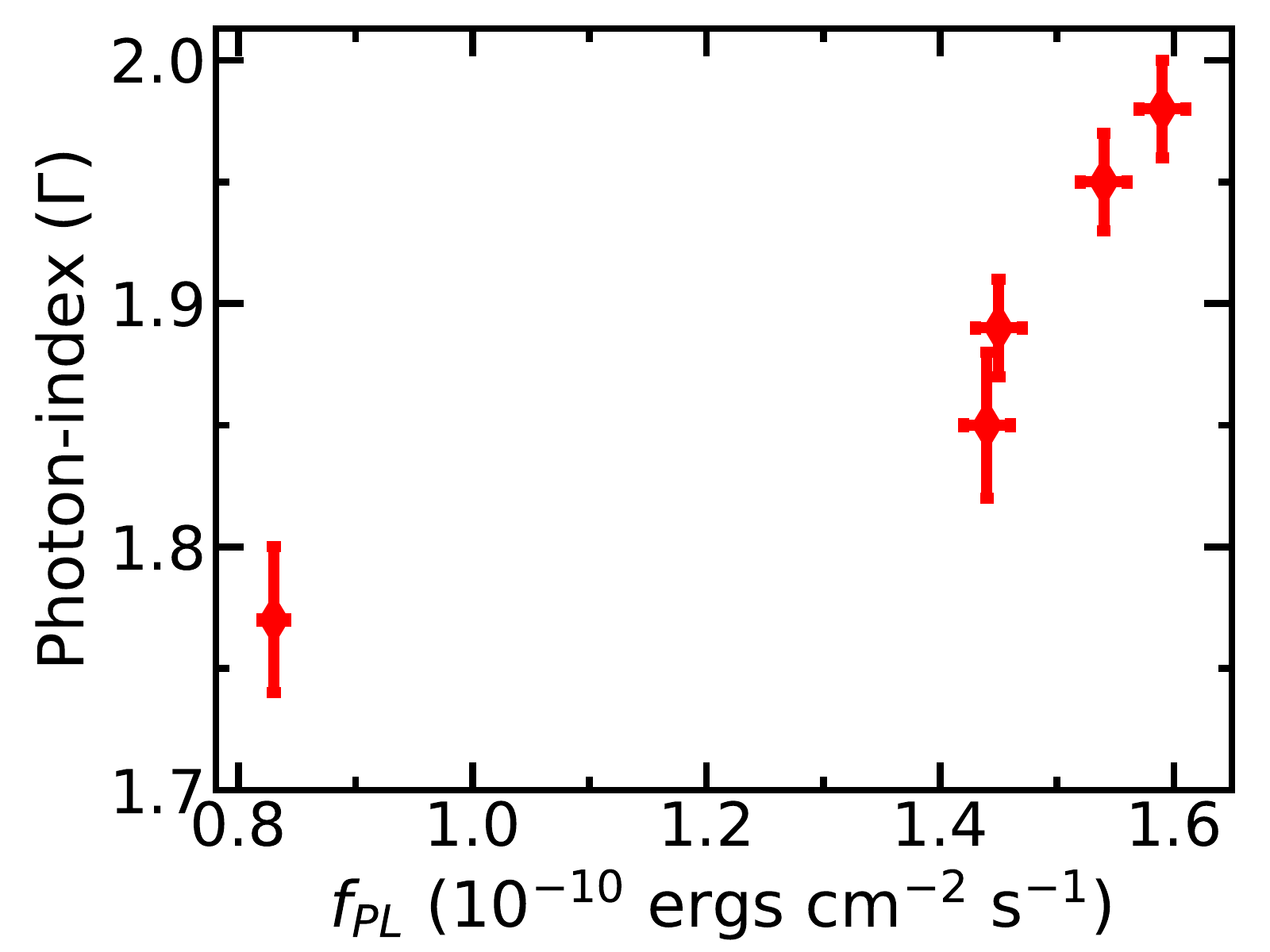}
 	\caption{The X-ray power-law photon index ($\Gamma$) versus the power-law flux, $f_{PL}$.}
 	\label{fig:gamma-vs-fpl}
 \end{figure}

Our analysis has revealed a variable X-ray spectrum, with the power-law  slope becoming steeper ($\Gamma\sim 1.8$ to $2$) with increasing X--ray flux ($f_{PL} \sim 8.3-15.9\times10^{-11}{\rm~ergs~cm^{-2}~s^{-1}}$; see Table~\ref{tab:xray_spec}). Figure~\ref{fig:gamma-vs-fpl} shows a plot of the best-fit $\Gamma$ versus the continuum flux, where the "softer when brighter" trend is apparent. 
Such spectral variability has been found in a number of Seyfert galaxies \citep{2002A&A...390...65D, 10.1046/j.1365-8711.2003.06556.x, 10.1111/j.1365-2966.2009.15382.x, 2011MNRAS.415.1895E, 10.1093/mnras/stu2571}, and is generally interpreted in terms of thermal Comptonization of seed optical/UV photons where the spectral steepening results from cooling of the coronal plasma as a result of increased seed flux 
\citep{Zdziarski_2001, 2004A&A...413..477P}.  

\begin{figure*}
 	\centering

 	\includegraphics[scale=0.35]{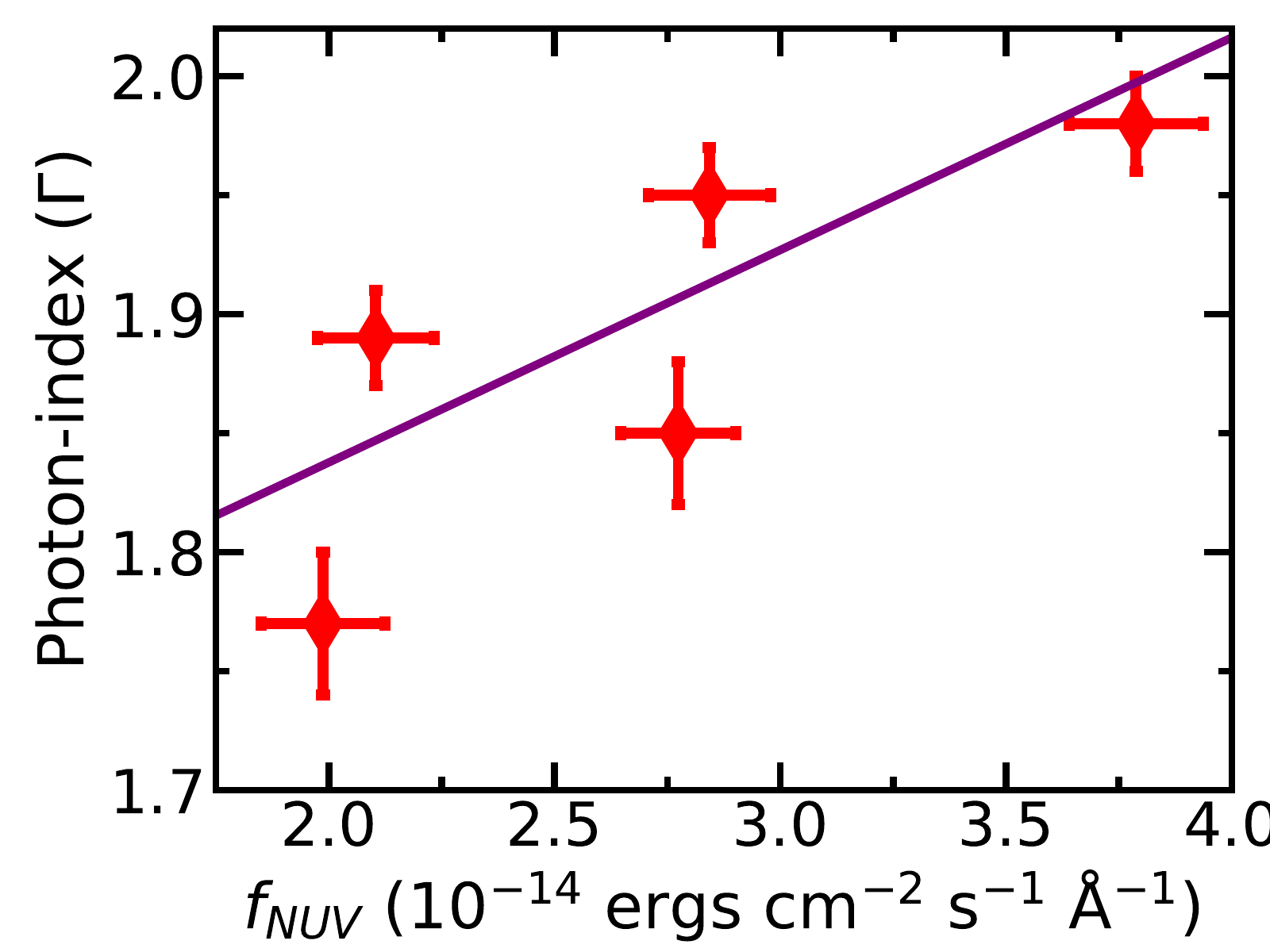}
	\includegraphics[scale=0.35]{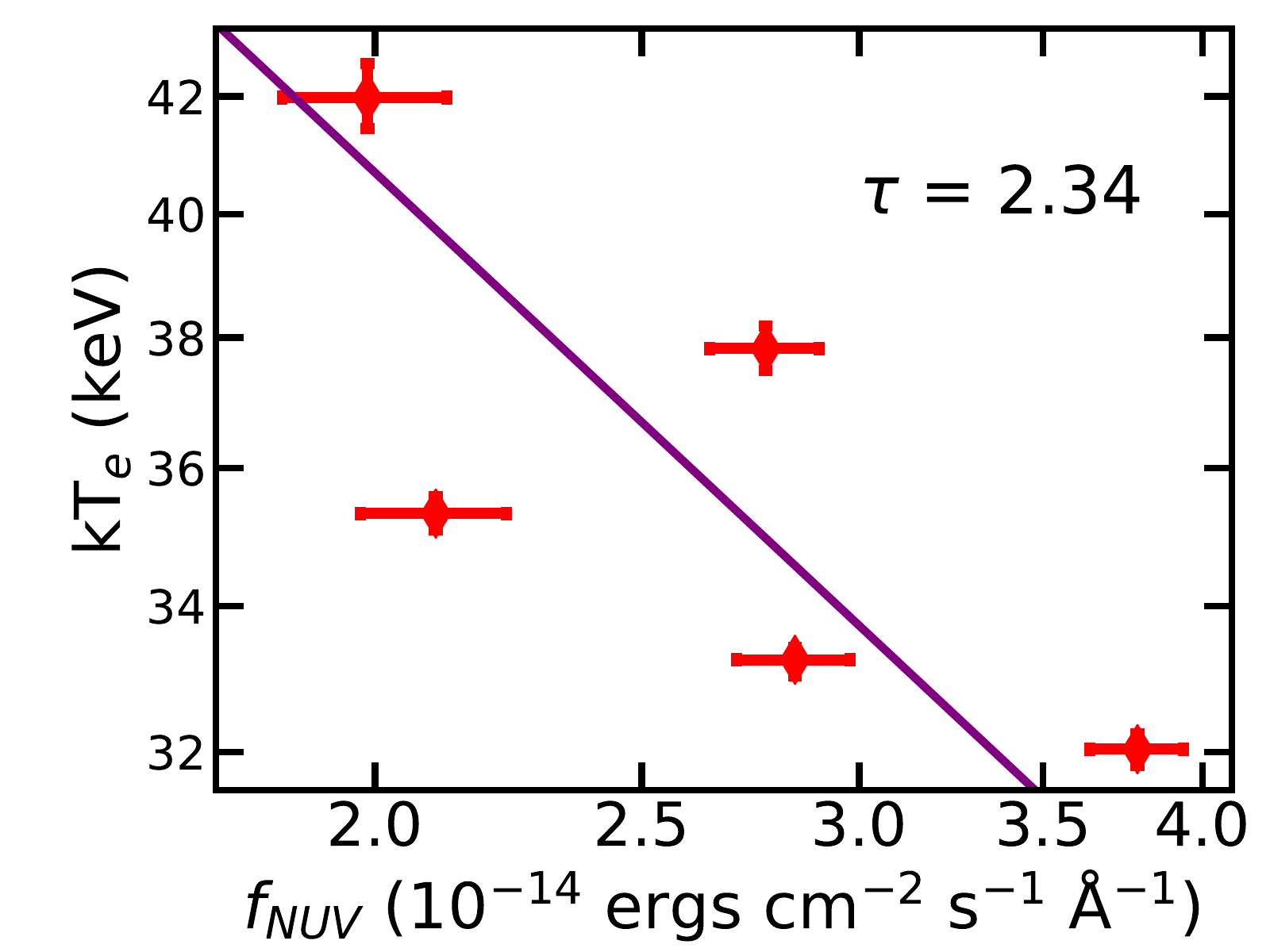}
	\includegraphics[scale=0.35]{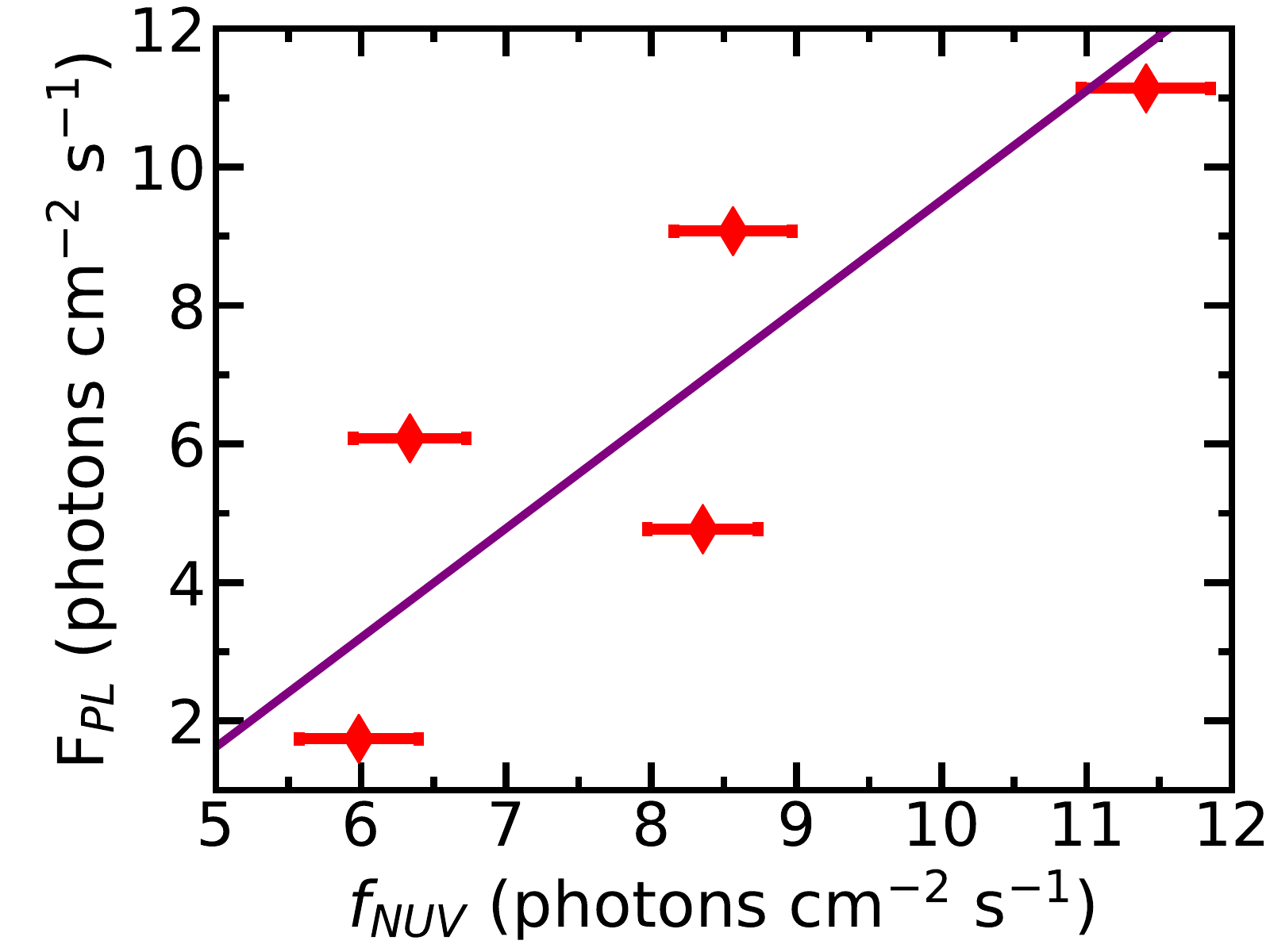} 		
 	\caption{ $\Gamma$ plotted as a function of the NUV flux, the corona temperature plotted as a function of the NUV flux, and the the broadband power-law photon flux in the 5\ev{}--500\kev{} band plotted against the intrinsic NUV photon flux (left, middle and right panels, respectively). Solid lines show the best-fit linear models to the data.}
 	\label{fig:spec_var}	
 \end{figure*}

Using the simultaneous UV/X-ray monitoring observations with \astrosat{}, we found that the X-ray spectral slope is positively correlated with the NUV flux, as well. This is clearly seen in the  left panel of Figure~\ref{fig:spec_var}. The solid line in this plot indicates the best linear-fit to the data (estimated taking into account the error on both variables). The best-fit slope is $8.9 \pm 3.9 \times 10^{12}$, which implies that the correlation is significant at more than the $2\sigma$ level. 

Consequently, the steepening of the X-ray photon-index with increasing NUV emission supports the hypothesis of thermal Comptonization for the X--ray emission in AGN. This is further supported by the plot in the middle panel of Figure~\ref{fig:spec_var}. This plot shows the best-fit coronal temperature, $kT_e$, as a function of the NUV intrinsic flux, in log-log space (the corona temperature values are those listed in Table~\ref{tab:thcomp} , when the combined NUV/X-ray spectra were fitted with the thermal Comptonization model at a constant optical depth). {The solid line shows the best-fit linear fit to the data (in log-log space; best-fitting slope of 0.5$\pm 0.2$)}. The decreasing temperature with increasing NUV flux clearly supports the scenario of seed photons cooling the hot corona. Of course, the scenario of a constant corona with a variable optical depth can also fit the data, as we already discussed in \S~\ref{broadbandspec}. But in this case, the $\tau$ variability anti-correlates with the NUV variations. We cannot think of a possible physical explanation for such a trend. 

If the variable seed flux from the accretion disk is the main driver of the X-ray spectral variability via thermal Comptonization, we would also expect a strong correlation between the NUV and X-ray power-law flux as the photon number must be conserved in the Comptonization process. This is already apparent from the normalized NUV and $f_{PL}$ light curves plotted in Figure~\ref{figintrinsicflux}. To investigate this issue further, we also calculated the X-ray power-law photon flux in the $5\ev$ to $500\kev$ band. To do so, we assumed a power-law model with an exponential cut-off. We used the best-fit power-law parameters listed in Table~\ref{tab:xray_spec} and the best-fit temperature values in Table~\ref{tab:thcomp} to compute the cut-off energy, $E_c$  (assuming $E_c = 3kT_e$). The right panel in Figure~\ref{fig:spec_var} shows the X-ray power-law
photon flux ($F_{PL}$) plotted against the intrinsic NUV photon flux. {Solid line shows again the best-fit linear relation (best-fit slope = $1.6\pm 0.5$)}. It is clear that the accretion disk emission and the X-ray power-law photon flux are strongly correlated, thus further supporting the idea that the accretion disk providing the seed photons for thermal Comptonization. The scatter seen in all plots in Figure~\ref{fig:spec_var}~({\it right}) could arise due to the effects of the intrinsic variability of the hot corona emission.

\subsection{The origin of the soft excess}

We detected a soft X-ray excess component below $\sim2\kev$, consistent with previous results \citep{2018A&A...619A..20M}. We did not detect an iron line in our spectra as the effective area of the SXT is small above $6\kev$ and the spectral resolution of the LAXPC  is poor. Also, the broad iron line is weak with an equivalent width of $\sim65\ev$ \citep{2007MNRAS.382..194N}. Therefore, the non-detection of a broad iron line in our analysis is consistent with the earlier detection of a weak, moderately broad iron line \citep{2004ApJ...608..157M, 2006ApJ...646..783M,2019ApJ...875..115O}. These results indicate that X--ray reflection from the inner disk is weak in this source.  \citet{2019ApJ...875..115O} fitted the $\sim 0.7-70\kev$ broadband \suzaku{}+\nustar{} spectral data with a relativistic reflection model and inferred a low reflection fraction ($R\sim 3.2\times10^{-3}$) from a truncated accretion disk ($R_{in}\sim 87R_g$). This result suggests that, at least some part of the soft excess observed in IC~4329A could be due to X--ray reflection from the disk at large radii. {However, we did not fit the data with a relativistic disk reflection component to investigate this issue in detail.} 


On the other hand, thermal Comptonization of the disk seed photons in a warm corona which is located above the inner disk could also produce the soft X-ray excess emission that we observe in many AGN (\citealt{2015A&A...575A..22M}, \citealt{2018A&A...611A..59P}, \citealt{2020A&A...634A..85P}). We actually found that the soft excess component in this object can be fitted well by the thermal Comptonization model {\textsc{nthcomp}}, with the best-fit warm corona temperature and spectral photon-index being $\sim 0.26\kev$ and $\leq 2.54$ ($3\sigma$ upper limit). Both these values are consistent with the values generally observed for type~I AGN \citep{2018MNRAS.480.1247K}.

\section{Conclusion}
\label{conclusion}
We analyzed high-resolution NUV images of IC~4329A and simultaneous X-ray data acquired with \astrosat{}, and found that the intrinsic UV emission from the AGN varied more than the X-ray emission. The intrinsically variable UV emission is correlated with both the X-ray continuum spectral slope and flux. Our observations are consistent with the accretion disk providing the seed photons for the thermal Comptonization process in the hot corona, and the observed X-ray spectral variability can be explained as the increasing UV emission cooling the hot corona at a constant optical depth. Future simultaneous broadband UV \& X-ray spectroscopic monitoring observations {with \astrosat{} and \nustar{}} could determine both the temperature and optical depth of the corona and the effect of changing seed flux on these characteristic coronal properties. Finally, we also detect a soft excess, which is well fitted by a black-body component with a temperature of $0.26\keV$, similar to what has been observed in other AGN as well. This component could be the result of a `warm' corona, located above the inner disc, which is up-scattering the disk photons emitted by the inner disk radius at $\sim 100 R_g$. {Perhaps some of the soft excess flux may also be due to the X-ray reflection from the inner disk, which in this source is either far from the center or filled with warm, optically thick medium \citep{Dewangan_et_al_2021}}.

\acknowledgments

{ We thank the anonymous referee for useful comments that improved the manuscript.}
This publication uses the data from the \astrosat{} mission of the Indian Space
Research Organisation (ISRO), archived at the Indian Space Science Data Centre
(ISSDC). This publication uses the data from the UVIT, SXT and LAXPC. The SXT and LAXPC data were processed by the pipeline software provided by the respective payload operation centers (POCs) at TIFR, Mumbai.
The UVIT data were checked and verified by the POC at IIA, Bangalore, and processed by the CCDLAB pipeline\cite{2017PASP..129k5002P}.
 This research has made use of the $python$ and $julia$ packages.
This research has made use of the SIMBAD/NED database.   PT acknowledges the University
Grant Commission (UGC), Government of India for financial supports.

\vspace{5mm}
\facilities{AstroSat}

\software{CCDLAB \citep{2017PASP..129k5002P},
			XSPEC \citep{1996ASPC..101...17A},
		Sherpa \citep{2001SPIE.4477...76F},
			SAOImageDS9 \citep{2003ASPC..295..489J},
			Julia \citep{doi:10.1137/141000671},
			Astropy \citep{2013A&A...558A..33A}
          }

\bibliography{mybib}
\bibliographystyle{aasjournal}

\end{document}